\documentclass[fleqn,usenatbib]{mnras}

\usepackage{newtxtext,newtxmath}
\usepackage[normalem]{ulem}

\usepackage[T1]{fontenc}

\DeclareRobustCommand{\VAN}[3]{#2}
\let\VANthebibliography\thebibliography
\def\thebibliography{\DeclareRobustCommand{\VAN}[3]{##3}\VANthebibliography}

\usepackage{graphicx}	
\usepackage{amsmath}	

\usepackage{diagbox}
\usepackage{makecell}
\usepackage[table]{xcolor}
\usepackage{pdflscape}

\def \aj {AJ}
\def \mnras {MNRAS}
\def \pasp {PASP}
\def \apj {ApJ}
\def \apjs {ApJS}
\def \apjl {ApJL}
\def \aap {A\&A}
\def \nat {Nature}
\def \araa {ARAA}

\def \aaps {A\&A Suppl.}
\def \pasa {PASA}

\title[The environment of iPTF13bvn]{An environmental analysis of supernova iPTF13bvn with HST and MUSE}

\author[Singleton et al.]{
Adam J. Singleton,$^{1}$\thanks{E-mail: AJSingleton1@sheffield.ac.uk (AJS)}, Justyn R. Maund$^{2}$, Ning-Chen Sun$^{3}$ 
\\
$^{1}$School of Mathematical and Physical Sciences, University of Sheffield, Hicks Building, Hounsfield Road, Sheffield S3 7RH, United Kingdom\\
$^{2}$Department of Physics, Royal Holloway, University of London, Egham Hill, Egham, TW20 0EX, United Kingdom\\
$^{3}$School of Astronomy and Space Science, University of Chinese Academy of Sciences, Beĳing 100049, People’s Republic of China\\
}

\date{Accepted XXX. Received YYY; in original form ZZZ}

\pubyear{2025}

\begin{document}
\label{firstpage}
\pagerange{\pageref{firstpage}--\pageref{lastpage}}
\maketitle

\begin{abstract}
Searches for supernovae (SNe) progenitors have relied on a direct detection of the star in fortuitous pre-explosion images. We propose an alternative method, using a combination of photometric stellar population fitting alongside integral-field-unit (IFU) spectroscopic analysis of the ionised gas to fully explore the SN environment and constrain the progenitor properties. Isochrone fitting of HST/WFC3 observations reveals the environment of iPTF13bvn contains two stellar populations with unique age ($\tau=\log{t\, \textrm{years}}$) and extinction ($A_V$) values, with the closest agreement found between past progenitor studies of iPTF13bvn and our oldest stellar population (P2): $\tau_{P2}=6.97^{+0.06}_{-0.06}$, a corresponding initial mass $M_{\textrm{initial},P2} = 20.0\, \textrm{M}_\odot$ and $A_{V,P2}=0.53^{+0.10}_{-0.08}\, \textrm{mag}$. Further analysis with VLT/MUSE IFU-spectroscopic observations reveals no bright H \textsc{ii} regions associated with iPTF13bvn, suggesting no immediate ongoing star formation. Extinctions derived from the ionised gas are a minimum of $\sim2.5$ times higher than the resolved stellar population values, assisting in building a 3D picture of the environment. An analysis of the distribution of spaxel extinctions reveals increased variability in the environment of iPTF13bvn, on the edge of a spiral arm. Our study highlights the complex relationship between stars, gas and dust and how, when used in a holistic environmental analysis, they can begin to resolve degeneracies that have plagued past progenitor investigations. Specifically for iPTF13bvn, our results support a binary progenitor and a growing consensus for binarity as the predominant mass-loss mechanism for Type Ib SNe progenitors.
\end{abstract}




\section{INTRODUCTION}
\label{sec:intro}

Massive stars with initial masses M$_\textrm{initial} \gtrsim 8M_\odot$ will end their lives as core-collapse supernovae (CCSNe) \citep{2004Sci...303..499S}. Compared to the thermonuclear detonation of white dwarfs observed as Type Ia supernovae (SNe), CCSNe have a variety of progenitors with diverse pre-explosion histories. The observational classifications of CCSNe are distinguished by spectral and lightcurve features, initially separated by the presence (Type II) or absence (Type Ib) of hydrogen. Further classifications are defined as follows: a lack of H and He lines (Type Ic); prominent and narrow H (Type IIn) or He (Type Ibn) lines generated by the interaction of the SN ejecta with dense circumstellar material (CSM); broad-lined spectral features (Type Ic-BL) due to high expansion velocities (for examples see \citealt{1997ARA&A..35..309F}). Bridging the gap between Type II and Type Ib, Type IIb SNe initially show signatures of H that quickly vanish and evolve to resemble a Type Ib. It is widely accepted that distinctions between these classifications are the product of a myriad of stellar evolution channels, greatly impacted by the initial mass of the progenitor and its mass-loss history \citep{2009MNRAS.395.1409S, 2013MNRAS.436..774E}. Mass loss occurs in massive stars through two main channels: strong stellar winds or binary interactions, with the former showing a strong metallicity dependence in single stars \citep{2003ApJ...591..288H}. During later phases of nuclear burning, however, massive stars can experience pulsations and eruptions as alternative forms of mass-loss \citep{2014ARA&A..52..487S}. Building up the ‘progenitor tree’ is crucial for disentangling whether all SNe are the result of the same explosion mechanism and how various factors (metallicity; initial mass; mass-loss; CSM; binarity; stellar environment) give rise to different explosion characteristics.

The ‘gold standard’ for progenitor detections requires deep pre-explosion archival images, typically sourced from the extensive Hubble Space Telescope (HST) data archive. Later, post-explosion images are required to conclusively claim that the progenitor candidate has disappeared from the SNe site. In the last two decades, the search for the progenitors of Type II SNe has converged on red supergiants (RSG) \citep{2004Sci...303..499S, 2004Natur.427..129M, 2009MNRAS.395.1409S, 2009ARA&A..47...63S}, which aligns well with theoretical predictions of stellar evolution \citep{1970AcA....20...47P}. SN 1993J \citep{1993Natur.364..507N, 1994AJ....107..662A, 1994ApJ...429..300W} and SN 2011dh \citep{2011ApJ...739L..37M, 2012ApJ...757...31B, 2015A&A...580A.142E} are examples of well studied Type IIb SNe with direct detections of their progenitors. The first was identified as a non-variable RSG of $13-20\, \textrm{M}_\odot$ and the second a 13 $\textrm{M}_\odot$ yellow supergiant, both with binary companions responsible for extensive stripping of their H envelopes prior to explosion. Detections of Type Ibc progenitors have remained the most elusive, with respect to Type II. iPTF13bvn was the first Type Ib SN with a possible progenitor detection, followed by Type Ic SN 2017ein \citep{2018MNRAS.480.2072K, 2018ApJ...860...90V, 2019ApJ...871..176X} and Type Ib SN 2019yvr \citep{2021MNRAS.504.2073K, 2022MNRAS.510.3701S}. These studies suggest an assortment of possible progenitors: RSGs, yellow supergiants (YSG), helium giants, luminous blue variables (LBVs), massive binaries (Wolf-Rayet or B stars) or cool and inflated stars in close binaries. Often the best-estimate properties of a progenitor produces a dichotomy in its proposed evolution, as it can not be determined whether the system is singular or binary. A growing consensus in the past decade suggests that binary interaction represents a significant (if not the predominant) mass-loss channel for the progenitors of stripped envelope supernovae (SESNe). The winds of single stars cannot independently reproduce the rates and diversity of SESNe \citep{2014ARA&A..52..487S, 2019sros.confE..73S}. Relying solely on the ‘gold standard’ method of constraining the properties of SNe progenitors yields few direct detections annually. Moreover, initially proposed progenitors can be false and later ruled out by late-time imaging \citep{2024arXiv241117969Z}. 

The first examples of environmental studies of SNe investigated correlations between the host galaxy morphology and the classifications of observed SNe (then only Type I vs. Type II). This found that CCSNe occur exclusively in late-time, star-forming galaxies \citep{1934PNAS...20..254B, 1953PASP...65..242R, 1959AnAp...22..123V}. With instrumental advancements and the expansion of the SNe classification scheme, later studies could begin to examine the local environments of CCSNe. \citet{1992AJ....103.1788V} studied the distances of SNe (now Type Ia, Type Ibc and Type II) explosion sites to the nearest H \textsc{ii} region, finding that CCSNe show a strong association with H \textsc{ii} regions and therefore high mass progenitors. Further studies \citep{1994PASP..106.1276B, 1996AJ....111.2017V} found similar results. In the 21st century, the introduction of larger SN (and extragalactic) catalogues prompted statistical analyses and, as \citet{2015PASA...32...19A} outline in a review, there has historically been three branches of analysis: 
\begin{enumerate}
	\item Host galaxy pixel statistics. 
	\item Radial distributions.
	\item H \textsc{ii} region metallicities. 
\end{enumerate}       
A popular approach to pixel statistics is Normalised Cumulative-Rank (NCR) method, used by \citet{2006A&A...453...57J} and \citeyear{2008MNRAS.390.1527A} to investigate the spatial association of SNe types with ongoing star formation (SF), using H$\alpha$ as a proxy, in their host galaxies. These studies probe median physical sizes on the order of 300 pc, meaning individual H \textsc{ii} regions become grouped into larger star forming regions. \citet{2006A&A...453...57J} is also an example of a radial distribution study, investigating where within their host galaxies different SNe explosion sites fall. H \textsc{ii} region metallicity studies \citep{2008IAUS..250..503M, 2010MNRAS.407.2660A, 2013AJ....146...31K, 2013AJ....146...30K, 2013A&A...558A.143T} generally find that Type Ibc SNe have higher metallicities than Type II (although this difference is small or undetected in some studies). The spectroscopic data in these studies suffers from a biased association to the nearest bright H \textsc{ii} region, as emission lines directly from the SNe sites are too weak.

Recent studies \citep{2016MNRAS.456.3175M, 2017MNRAS.469.2202M, 2021MNRAS.504.2253S, 2022MNRAS.512.1541G, 2023MNRAS.521.2860S} considered individual stars or ionised gas in the SN environment to constrain the progenitor age, extinction and metallicity. \cite{2016MNRAS.456.3175M} used a Bayesian approach with Nested Sampling to fit the surrounding stellar populations, providing age and extinction estimates of 23 stripped-envelope SNe (Type IIb \& Type Ibc). Their characteristic ages support a substantial number of high mass ($M_{initial} > 30 \textrm{M}_\odot$) progenitors for Type Ibc SNe. \cite{2021MNRAS.504.2253S} expanded upon this analysis by including integral-field-unit (IFU) spectroscopy to model the ionised gas component of two SNe environments (2004dg[II-P] \& 2012P[IIb]), revealing a star forming complex that has photo-ionised a giant H \textsc{ii} region within NGC 5806. \citet{2022MNRAS.510.3701S} repeated a similar analysis for the Type Ib SN 2019yvr. In this paper we will adapt this approach to studying the environments of CCSNe, utilising the high spatial information of HST and MUSE data to combat the limitations of past studies. 

The SN iPTF13bvn was discovered by \citet{2013ApJ...775L...7C} and reported by the intermediate Palomar Transient Factory (iPTF) on 2013 June 16.238 UT, in the galaxy NGC 5806. An early-time spectral and photometric series classified iPTF13bvn as a Type Ib SN. \citet{2013ApJ...775L...7C} observed iPTF13bvn at radio wavlengths; synchrotron self-absorption modelling produced mass-loss parameters consistent with a Wolf-Rayet progenitor. Conversely, \cite{2014AJ....148...68B} were the first to claim an interacting binary progenitor for iPTF13bvn through bolometric lightcurve fitting to hydrodynamic models. Subsequent studies have formed a growing consensus for the binary progenitor \citep{2015MNRAS.446.2689E, 2015AA...579A..95K, 2016MNRAS.461L.117E}. Errors in the initially quoted progenitor photometry for iPTF13bvn were identified and corrected by \cite{2015MNRAS.446.2689E}. Their reanalysis agreed with a range of binary models, but ruled out typical Wolf-Rayet models. A late-time study by \cite{2016MNRAS.461L.117E} confirmed that the identified progenitor had disappeared, as well as providing constraints on the suspected companion. 

This paper is outlined as follows. First, we describe the methods of processing and analysing our photometric and spectroscopic data (Section \ref{sec:methods}). This includes our Bayesian approach to isochrone fitting for the stars in the environment of iPTF13bvn (Sections \ref{sec:astrom} and \ref{sec:Bayes}), as well as our spectroscopic fitting algorithm to study the gas (Sections \ref{sec:MUSE} and \ref{sec:algorithm}). A shorter, intersectional study between both datasets is described in Section \ref{sec:resample}. The results of all parts of our environmental analysis are then presented separately in Section \ref{sec:results}. Finally the implications, limitations and context of our results, with respect to past studies of iPTF13bvn, are discussed in Section \ref{sec:discussion}, followed by our conclusions and future work (Section \ref{sec:conclusion}).      

For this study we use the foreground Galactic extinction in the $V$-band \citep{2011ApJ...737..103S}, quoted from the NASA/IPAC Extragalactic Database (NED)\footnote{https://ned.ipac.caltech.edu/} towards NGC 5806 of $A_V = 0.139\, \textrm{mag}$. We adopt the Tully-Fisher distance to NGC 5806 of 22.5 Mpc \citep{2009AJ....138..323T}, corresponding to a distance modulus $\mu=31.76 \pm 0.36\, \textrm{mag}$ for consistency with \citet{2015MNRAS.446.2689E}.

\section{OBSERVATIONS AND METHODS}
\label{sec:methods}

Our analysis of the environment around iPTF13bvn consists of two components: 1) a photometric study of resolved stars in the environment that relies on long-exposure, high-spatial-resolution images of the host galaxy; and 2) a comparative study of the gas/dust/stars across a broader environment of iPTF13bvn, using IFU spectroscopic data.

\begin{figure}
\centering
\includegraphics[width=85mm]{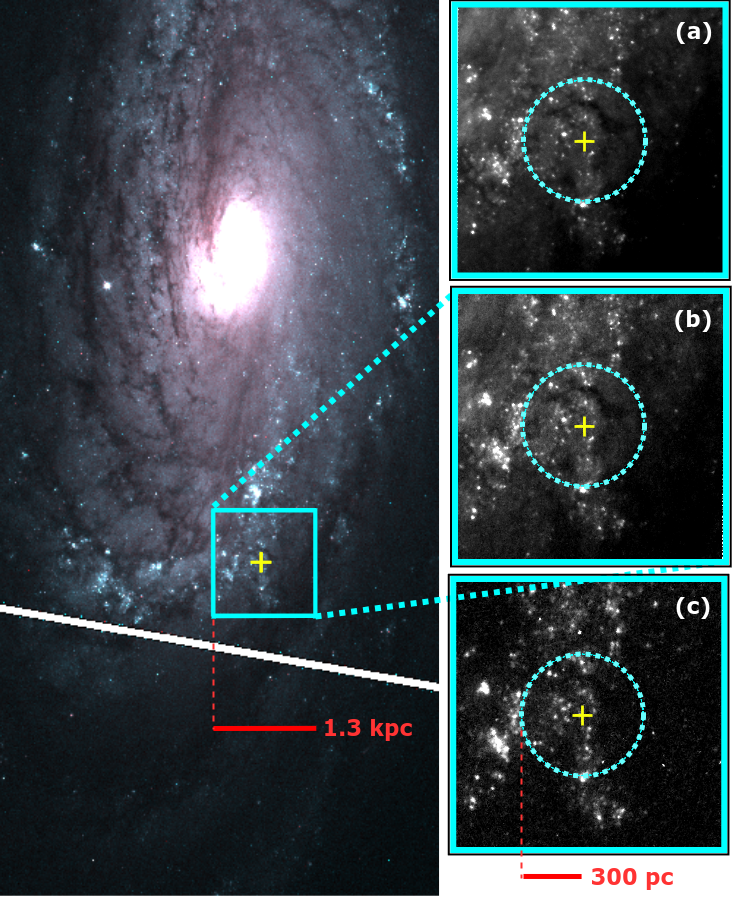}
\vspace{-2mm}
\caption{Left: A false colour HST image of NGC 5806 using the F555W and F438W images. A cyan box highlights the extracted region of the VLT/MUSE IFU datacube, a gold cross marks the location of iPTF13bvn. Right: a close-up of MUSE region in all three HST filters, (a) F555W (b) F438W (c) F225W. A dotted cyan circle (radius = 300 pc) highlights the environment definition used to resolve stellar populations.}
\label{fig:environment}
\end{figure}

\subsection{HST/WFC3}
\label{sec:HST} 

\subsubsection{Astrometry and Photometry}
\label{sec:astrom}

Hubble Space Telescope (HST) archival data for NGC~5806 were retrieved from the Mikulski Archive for Space Telescopes \footnote{\url{https://archive.stsci.edu/missions-and-data/hst}}, as detailed in Table \ref{tab:1}. Three long-exposure observations taken with the Wide Field Camera 3 (WFC3) with the Ultraviolet-Visible (UVIS) channel were obtained with the F225W (30 June 2015), F438W and F555W (25 June 2019) wide-band filters. The wavelength coverage spans the visible through UV. WFC3/UVIS has a pixel scale of 0.04 arcsec and a $160\times160\, \textrm{arcsec}^2$ field of view (FoV). This dataset contains most of NGC 5806 (see Figure \ref{fig:environment}) and no residual light from iPTF13bvn. An additional WFC3/UVIS F555W (2 September 2013) subarray image of the site of iPTF13bvn, centred on the SN, was obtained for localising the SN position. Each download contains the raw files after charge transfer efficiency corrections (\texttt{*\_flc.fits}) and the combined file corrected for cosmic-rays and distortion (\texttt{\_drc.fits}) using the \textsc{astrodrizzle} package. The world coordinate system (WCS) of the three (\texttt{\_drc.fits}) files were aligned using \textsc{tweakreg} and \textsc{tweakback} from the \textsc{drizzlepac} package, with the F438W image used as the reference for the WCS.  

Point-spread-function (PSF) photometry was then performed using \textsc{dolphot} \citep{2000PASP..112.1383D} with the recommended input parameters and the F555W image (2019) used as the reference image, since this resulted in the largest number of common objects between all three WFC3 images. Resolved objects were filtered using \textsc{dolphot} PSF output parameters according to the following criteria:
\begin{enumerate}
    \item[(1)] type of source, TYPE $=1$ (corresponding to ‘star’);
    \item[(2)]photometry quality flag, FLAG $\leq3$;
    \item[(3)]source crowding, CROWD $\leq2$;
    \item[(4)]source sharpness, $-0.5\leq$ SHARP $\leq0.5$;
    \item[(5)]signal-to-noise ratio, SNR $\geq5$.
\end{enumerate}
An object was retained in the final photometric catalogue if it met these quality checks in a minimum of one WFC3 image. When an object did not meet the above criteria in all three images, non-detections were substituted with a detection limit and corresponding error. Detection limits were found using artificial star tests (AST) within \textsc{dolphot}. For each discrete magnitude value spanning a $22-33\, \textrm{mag}$ range, in increments of $0.5\, \textrm{mag}$, 15 artificial stars were inserted into the image and the detection threshold was assessed by the success rate of the recovery of the artificial stars. The number of recovered artificial stars across this range was fit with a cumulative normal distribution function, and the corresponding ‘50\% recovered’ magnitude was taken to be the detection limit. Non-linear least squares regression optimised the error function parameters, providing a standard deviation about the detection limit, and values for the detection limit in each HST/WFC3 image are listed in Table \ref{tab:1}.  

\begin{table*}
\caption{HST/WFC3 and VLT/MUSE data for the site of iPTF13bvn used in this study.}
    \begin{centering} 
    \begin{tabular}{c c c c c c c} 
    \hline
    \vtop{\hbox{\strut Program}\hbox{\strut ID}} & Date (UT) & Instrument & Filter/Mode & \vtop{\hbox{\strut Exposure}\hbox{\strut Time (s)}} & $m_j^{lim}$ (mag) & $\sigma_j^{lim}$ (mag) \\ [0.5ex] 
    \hline\hline
    $15152^{\mathrm{a}}$ & 2019-06-25 & HST/WFC3/UVIS & F555W & 5437.00 & 27.96 & +0.56 \\ 
    $15152^{\mathrm{a}}$ & 2019-06-25 & HST/WFC3/UVIS & F438W & 5408.00 & 27.78 & +0.58 \\
    $13822^{\mathrm{b}}$ & 2015-06-30 & HST/WFC3/UVIS & F225W & 8865.00 & 25.95 & +0.52 \\ 
    $12888^{\mathrm{a}}$ & 2013-09-02 & HST/WFC3/UVIS & F555W & 300.00 & - & - \\ [1ex] 
    $097$.B-$0165^{\mathrm{c}}$ & 2016-04-07 & VLT/MUSE/LIFU & WFM & 3600.00 & - & - \\ [1ex] 
    \end{tabular}
    \end{centering}
    \label{tab:1}
    \\
    $^{a}$ PI: S. D. Van Dyk\\
    $^{b}$ PI: G. Folatelli\\
    $^{c}$ PI: D. M. Carollo\\
\end{table*}
A post-explosion HST image of iPTF13bvn, acquired in 2013 (see Table \ref{tab:1}), was used to locate the SN on the later HST images. Photometry of the 2013 HST WFC3/UVIS F555W image, with \textsc{dolphot}, was used to derive the positions of 25 common stars, that were used to find a coordinate transformation to determine the location of the SN on the 2019 HST WFC3/UVIS F438W image. The location of iPTF13bvn was transformed with a precision of $(\pm0.18, \pm0.26)$ HST/WFC3 pixels. The stellar environment of iPTF13bvn was defined as a circular region of 300 pc, centred on the SN position. The 300 pc region (corresponding to 68.761 F555W image pixels) is represented by the dotted cyan circle in Figure \ref{fig:environment}, with the environment displayed in each HST/WFC3 filter long-exposure image on the right panel.

\subsubsection{Bayesian Approach to Stellar Population Fitting with Nested Sampling}
\label{sec:Bayes}

The Bayesian approach for determining the parameters ($\mathbf{x}$; e.g. age, extinction, etc.) of a stellar environment from a given set of data ($\mathbf{D}$) can be considered using Bayes’ theorem as:
\begin{equation}
    p(\mathbf{x}|\mathbf{D})= \frac{p(\mathbf{D}|\mathbf{x})p(\mathbf{x})}{p(\mathbf{D})}\ .    
\end{equation}
The posterior probability distribution, $p(\mathbf{x}|\mathbf{D})$, is comprised of the likelihood function, $p(\mathbf{D}|\mathbf{x})$, prior probability, $p(\mathbf{x})$, and evidence, $p(\mathbf{D})$. Following the methods employed by \cite{2005A&A...436..127J}, \cite{2016MNRAS.456.3175M} and \cite{2021MNRAS.504.2253S}, we use a hierarchical Bayesian mixture model for estimating the age ($\tau = \log_{10}{t\, \textrm{years}}$) and extinction ($A_V$) of $N_m$ stellar populations. The 'mixture model' allows us to build our model out of underlying subpopulations (i.e. multiple stellar population ages; single star and binary contributions), where final probabilities are taken as the sum of weighted probabilities from these subpopulations. For the $i^{th}$ star in a data set, $D_i$, the distance modulus, ${\mu}^{'}$, and stellar mass, $M_i$, are marginalised over as nuisance parameters, since any observed data from individual stars could suggest a range of masses. Assuming these parameters are independent, the prior probability for the stellar mass follows a \citet{1955ApJ...121..161S} Initial Mass Function (IMF) normalised over the mass range of our stellar evolution model and with an IMF exponent of $-2.35$. A \citet{1955ApJ...121..161S} IMF is sufficient, as we are handling massive stars in star-forming regions. In this case, we use the \textsc{parsec} stellar isochrones \citep{2002A&A...391..195G} as our stellar evolution model with the metallicity fixed at solar (Section \ref{sec:VLTenv}). 
For each star the likelihood function is a product of normal distributions which directly compare the observed apparent magnitude in the $j^{th}$ filter, $m_j$ (with error $\sigma_j$), to the predicted absolute magnitude of a star with mass $M_i$ on an isochrone of age $\tau$ and extinction $A_v$, $m_{j}(\tau, A_V, M_i)$, taking the form: 

\begin{equation}
\label{eqn:likelihood}
\resizebox{\columnwidth}{!}{
$
p(D_i \vert \tau,A_V,{\mu}^{'},M_i)=\underset{j}{\prod} \frac{1}{\sqrt{2\pi}\sigma_j}\exp\left\{-\frac{1}{2}\left(\frac{m_j-m_{j}(\tau,A_V,M_i)-{\mu}^{'}}{\sigma_j}\right)^2\right\}.
$
}
\end{equation}

The exception to this is a substitution for stars missing photometry in a specific filter. In this case the detection limits ($m_j^{lim}$, $\sigma_j^{lim}$) of the filter are characterised by a cumulative normal distribution, giving:

\begin{equation}
\label{eqn:sub}
\resizebox{\columnwidth}{!}{
$
p(m_j^{lim},\sigma_j^{lim} \vert \tau,A_V,{\mu}^{'},M_i)=\frac{1}{2}\left\{1+\mathrm{erf} \left(\frac{m_j^{lim}-m_j(\tau,A_V,M_i)-{\mu}^{'}}{\sqrt{2}\sigma_j^{lim}}\right)\right\}.
$
}
\end{equation}

This allows us to include multiple filters in our analysis even when the photometry is partially complete. Flat priors are used for both $\tau$ and $A_V$ and a normalisation condition is implemented on the likelihood function, such that it integrates to unity over the prior ranges of $\tau$ and $A_V$. At this stage in Equations \ref{eqn:likelihood} and \ref{eqn:sub}, $\tau$ and $A_V$ represent the predicted age and extinction of individual stars \textit{not} the overall stellar population estimates.

Implementation of the mixture model allows us to simultaneously include contributions from non-interacting binaries, as well as a mixture of $N_m$ populations (temporally separated epochs of star formation). Probability distributions are treated as the sum of the single and binary case weighted by the binary fraction $P_{bin}$, where $P_{bin} = 0.5$ is a fixed value throughout our analysis. A hierarchical approach uses each star separately to probe the distributions within the parameter space, and then uses the collection of stellar properties to fit a distribution to the entire population. We assume a normal distribution for the age and extinction when fitting for the collection of stars inside the selected environment to produce estimates for the overall stellar populations. For a complete formulation of this Bayesian approach and other considerations, the reader is directed to \citet{2016MNRAS.456.3175M}. The determination of the likelihood functions for each star uses the same approach further described in \citet{2021MNRAS.504.2253S}. 

In order to determine the Bayesian evidence (or marginal likelihood; $Z$) and posterior distribution from our predetermined likelihoods, we apply the Nested Sampling algorithm \citep{2004AIPC..735..395S} through the open-source code \textsc{ultranest} \citep{2021JOSS....6.3001B}. \cite{2016MNRAS.456.3175M} discuss two important considerations when using Nested Sampling, which are:
\begin{enumerate}
    \item[(i)]{Due to the interchangeable nature of the distribution labels in our model, the evidence is overestimated by a factor of $N_{m}!$ which can easily be corrected for. This means we allow the Nested Sampling algorithm to explore all modes of our mixture model, including mirror modes, requiring the results to then be sorted into $N_m$ components.} 
    \item[(ii)]{We use the Jeffreys’ scale \citep{1939thpr.book.....J} to inform how $Z$ indicates the number of age components that provide the best-fit model to the data. An increase in the corrected evidence ($\ln{[Z/N_{m}!]}$) greater than $\sim 30$ is deemed ‘strong evidence’ in support of that model.}
\end{enumerate}

\subsection{VLT/MUSE}
\label{sec:VLT}

\subsubsection{Datacube Analysis}
\label{sec:MUSE}

For the second phase of our environmental analysis, spatially-resolved integral-field-unit (IFU) spectroscopic data is required. The Multi-Unit Spectroscopic Explorer (MUSE) mounted on the Very Large Telescope (VLT) acquired IFU spectroscopy of NGC 5806 on 7 April 2016 (see Table \ref{tab:1}). Operating in the wide-field mode (WFM) provides a $1\times1\, \textrm{arcmin}^2$ FoV with $0.2\, \textrm{arcsec}$ spatial sampling covering a spectral range of 4750-9350 \r{A} with 1.25 \r{A} sampling. Once processed through the standard pipeline, the IFU datacube is publicly available through the European Southern Observatory (ESO) Archive Science Portal\footnote{\url{http://archive.eso.org}}. Manipulation of the datacube was achieved with the \textsc{mpdaf} \citep{2019ASPC..521..545P} and \textsc{ifuanal} \citep{2018MNRAS.473.1359L} packages. Following the alignment procedure presented in Section \ref{sec:astrom}, 8 common objects between the MUSE datacube and WFC3 images produced transformed coordinates for iPTF13bvn in the MUSE field. To improve the signal-to-noise ratio (SNR) of spectra across the datacube, $2\times2$ spaxels were resampled into single spaxels and their errors added in quadrature. This approach was chosen over the commonly used Voronoi tessellation approach for resampling IFU data, where spaxels are grouped into tiles of varying shape and size to ensure uniform spectral signal-to-noise ratios (SNRs). Although spectral SNR is still an important consideration within this work, it was more valuable for us to retain the spatial information within the environment of iPTF13bvn. A $30\times30$ spaxel (along the spatial axes) cube, centred on transformed coordinates of iPTF13bvn, was extracted and is highlighted by the cyan squares in Figure \ref{fig:environment}. Within this extracted cube the exact position of iPTF13bvn is at the binned spaxel position $(14.17\pm0.08, 14.19\pm0.11)$. Given that $(14,14)$ occurs at the centre of the spaxel, we can confidently conclude that iPTF13bvn falls inside this spaxel. Hereafter, references to the MUSE datacube or individual spaxels refer to this new, resampled, $30 \times 30$ spaxel datacube. The original, full FoV datacube of NGC 5806 can be found in Figures 1, 2, 3 and 4 of \cite{2021MNRAS.504.2253S}.

With \textsc{ifuanal}, the datacube was de-redshifted and dereddened assuming values of $z = 0.00449$ and $E(B-V) = 0.044\, \textrm{mag}$ (assuming a Galactic \citealt{1989ApJ...345..245C} reddening law with $R_V=3.1$). The galactic rotational velocity across this environment was found using a summed spectrum of the entire datacube, and a single Gaussian fitting to the Balmer $\textrm{H}\alpha$ emission line. Subtraction from the rest-frame peak wavelength gave an averaged rotational velocity of $v_{\textrm{rot}}=70.6\, \textrm{km/s}$. This estimate for our extracted region agrees with the corresponding values in the rotational velocity map of NGC 5806 by \cite{2021MNRAS.504.2253S}. Differential rotation across the extracted MUSE datacube is minimal, with the maximum discrepancy to our average $v_{\textrm{rot}}$ value corresponding to a $\Delta\lambda=2.5$ \r{A} (i.e. two spectral bins). A description of how we account for these minor central wavelength shifts is provided in Appendix \ref{sec:appendix}.     

\subsubsection{Continuum and Emission Line Fitting Algorithm}
\label{sec:algorithm}

Strong nebular lines within the dataset include the Balmer hydrogen recombination lines ($\textrm{H}\alpha$ $\lambda6563$ and $\textrm{H}\beta$ $\lambda4861$) and the collisionally-excited $\textrm{[N II]}$ $\lambda\lambda6548, 6584$ doublet. As expected, the $\lambda6548$ line is much weaker than the $\lambda6584$ line. Other weak lines that appear include: $\textrm{He I}$ $\lambda5876$, $\textrm{He I}$ $\lambda6678$, $\textrm{[O III]}$ $\lambda\lambda4959, 5007$, $\textrm{[O I]}$ $\lambda\lambda6300, 6363$ and $\textrm{[S II]}$ $\lambda\lambda6717, 6731$. Since the flux in our $2\times2$ binned spaxels can reach relatively low levels (mean continuum flux $= 54 \times 10^{-20}\,\mathrm{erg} \mathring{A}^{-1}\,\textrm{s}^{-1}\,\textrm{cm}^{-2}$), we focus on the $\textrm{H}\alpha$ $\lambda6563$ and $\textrm{H}\beta$ $\lambda4861$ lines in this study. 

Due to the relatively low spectral resolution of the MUSE data, an elementary approach to fitting was favoured. Function fitting to continua or nebular lines was performed using optimize.curve\_fit() in the \textsc{scipy} package and subsequent reduced-chi-squared ($\chi^{2}_r$) comparisons. All 900 spectra in the datacube had their continua fit with a third order polynomial spanning the full 4750-9350 \r{A} wavelength range, and a mean value taken to represent the continuum flux level of each spectrum. Narrow wavelength regions were extracted before fitting emission lines: 6520-6640 \r{A} and 4780-4940 \r{A} for $\textrm{H}\alpha$ and $\textrm{H}\beta$ respectively. Once the lines were masked, including the $\textrm{[N II]}$ lines either side of $\textrm{H}\alpha$, the remaining data points were linearly fit and continuum subtracted. The remaining emission was then fit with a series of functions:
\begin{description}
    \item[\textbf{(1) ‘Weak’ Double:}] weak signal spectra fit with a double Gaussian, accounting for both emission and absorption. 
    \item[\textbf{(2) ‘Strong’ Double:}] strong signal spectra fit with a double Gaussian, accounting for both emission and absorption.
    \item[\textbf{(3) ‘Weak’ Single:}] weak signal spectra fit with a single Gaussian, accounting solely for emission.
    \item[\textbf{(4) ‘Strong’ Single:}] strong signal spectra fit with a single Gaussian, accounting solely for emission.
    \item[\textbf{(5) Absorption:}] single Gaussian fit to pure absorption feature. 
    \item[\textbf{(6) Continuum:}] linear fit to featureless spectra. 
\end{description}
Here, distinctions between 'Weak' and 'Strong' correspond to different starting values for the optimisation of the Gaussian curves: initial emission amplitudes are $\times12$ larger for \textbf{(2)} and \textbf{(4)} compared to \textbf{(1)} and \textbf{(3)}; initial absorption amplitudes are $\times2.5$ larger for \textbf{(2)} compared to \textbf{(1)}. Out of these six options, the corresponding minimum $\chi^{2}_r$ is subjected to a series of comparisons before being selected as the optimum fitting. For example, the difference between the minimum $\chi^{2}_r$ and the linear continuum \textbf{(6)} $\chi^{2}_r$ must surpass 100 to maintain ‘optimum’ status. Other $\chi^{2}_r$ comparisons between ‘Double’ and ‘Single’ Gaussian fits are included to avoid overfitting of the lowest flux emission lines, and a discussion surrounding the choice of a Gaussian fit to any absoprtion features is provided in Appendix \ref{sec:appendix}. A SNR $\geq 15$ is required for any emission line to be considered for a double Gaussian fitting. Spectral features are only considered emission if the peak exceeds 3$\sigma$ above the mean continuum flux level. The series of $\chi^{2}_r$ comparison criteria were deduced through trial and error; although an extensive list is not provided, further conditions were included to minimise overfitting to spectral noise or remaining artifacts from the MUSE data pipeline. Post-filtering, the surviving $\chi^{2}_r$ for fitting types \textbf{(1)} through \textbf{(6)} is labelled as the ‘best fitting’ and used to derive physical values: integrated flux, full width at half maximum (FWHM), equivalent width (EW) and V-band extinction ($A_V$). Emission flux errors were estimated by performing 1000 random draws from a multivariate normal distribution of the optimised Gaussian parameters and corresponding covariance matrix. Each randomly drawn Gaussian is integrated and the standard deviation of the resulting sample is taken as the final flux error. Some examples of $\textrm{H}\alpha$ and $\textrm{H}\beta$ line fitting are given in Appendix \ref{sec:appendix}. With this overall approach to spectral fitting, we are not fitting for the underlying stellar spectral energy distribution (SED) as is standard when running \textsc{starlight} under the \textsc{ifuanal} package \citep{2018MNRAS.473.1359L}. Instead, we extract pure emission features by accounting for detectable absorption features in individual spaxel spectra. Our chain of comparisons for the reduced-chi-squared fitting metric avoids fitting to non-physical spectral features and handles statistical noise within each spectrum. We are, however, unable to predict any remaining features of the data reduction pipeline that may plague some spectra.        

Ionised gas extinctions were found through the Balmer decrement method, utilising a set of intrinsic flux ratios ($I[\textrm{H}\alpha]/I[\textrm{H}\beta] = 3.42, 3.10, 2.86 \textrm{ and } 2.69$) from \citet{2006agna.book.....O}, with corresponding gas temperatures (T$ = 2500, 5000, 10000 \textrm{ and } 20000 \textrm{ K}$ respectively). Conversions to V-band extinctions ($A_V$) used the \citealt{1989ApJ...345..245C} extinction law with $R_{V} = A_V/E(B-V) = 3.1$. 

\subsection{Resampling HST-to-MUSE}
\label{sec:resample}

For an intersectional study between the two datasets, we resample the HST images into maps of surface brightness on the same spatial pixel scale as MUSE. We follow the guidelines on (\url{https://www.stsci.edu/hst/instrumentation/wfc3/data-analysis/photometric-calibration/uvis-encircled-energy}) to convert pixel count rates into Vega magnitudes with 
\begin{equation}
m_{\textrm{Vega}} = \textrm{zeropoint} - 2.5\log{(\textrm{counts} - \textrm{background})} - 2.5\log{(1/\textrm{EE})}. 
\label{eqn:convert}
\end{equation}
Values substituted for zeropoints and encircled energy (EE) corrections are listed in Table \ref{tab:2} for our three main HST images. All EE corrections are for an aperture of radius $= 0.20\, \textrm{arcseconds}$, equivalent to a single MUSE spaxel. Also in Table \ref{tab:2} are newly found background estimates for our F555W, F438W and F225W images. Processed images from the HST archives are background subtracted with a single value, corresponding to a peak in a histogram of dark pixels. Our new background estimates were found by averaging across dark pixels; their low ($10^{-3}$) negative values could imply a slight oversubtraction of the background in the HST/WFC3 pipeline.   

\begin{table}
\caption{Values substituted into Equation \ref{eqn:convert} to convert HST image electron counts into surface brightness magnitudes: backgrounds, zeropoints and Encircled Energy (EE) corrections.}
    \centering 
    \begin{tabular}{c c c c} 
    \hline
    Filter & \vtop{\hbox{\strut Background}\hbox{\strut $10^{-3}$ (e$^{-}$/s)}} & \vtop{\hbox{\strut Zeropoint}\hbox{\strut (mag)}} & \vtop{\hbox{\strut EE Corrections}\hbox{\strut [r = 0.20”]}} \\ [0.5ex] 
    \hline\hline
    F555W & -8.95 & 25.823 & 0.85435 \\ 
    \hline
    F438W & -5.54 & 24.990 & 0.85247 \\
    \hline
    F225W & -2.76 & 22.646 & 0.78 \\ 
    \hline
    \end{tabular}
    \label{tab:2}
\end{table}

\section{RESULTS}
\label{sec:results}

\subsection{Stellar Populations and the Environment of iPTF13bvn as seen through HST/WFC3}
\label{sec:bayesresults}

As can be seen in Figure \ref{fig:environment}, iPTF13bvn is located in one of the spiral arms of NGC 5806 where massive star formation would be expected. iPTF13bvn appears to be situated on the edge of the spiral arm, such that the west portion of the environment is less populated. In this lower density region, a dark filament feature appears in all three of the WFC3 filters that wraps around the site of iPTF13bvn and could imply a gas/dust structure along the line-of-sight. Performing PSF photometry with \textsc{dolphot} yields 138 ‘good quality’ stars in the environment of iPTF13bvn (cyan dashed circle in Figure \ref{fig:environment}). The majority of magnitude detection limit substitutions occur for the F225W filter (93 substitutions), followed by F438W (54 substitutions) and minimal for F555W (4 substitutions). Given that our definition of the environment ($r=300$ pc) encloses a relatively large volume when considering regions of star formation, 138 resolved stars implies a sparse population of massive stars.

When applying our Bayesian approach the $k^{\textrm{th}}$ stellar population has three free parameters: $\log{(\textrm{age years})}$ ($\tau_{k}$), extinction ($A_{V,k}$) and extinction dispersion ($\delta\textrm{A}_\textrm{V,k}$) which relates to the standard deviation of the normal distribution through $\sigma_{A_{V,k}}=0.05\times10^{\delta A_{V,k}}$. Likelihood functions were obtained for each of the 138 stars, following the formulation described in Section \ref{sec:Bayes}. \textsc{parsec} isochrones were downloaded over the logarithmic age range of $6.0 \leq \tau \leq 8.0$ in steps of 0.01 and a fixed metallicity fraction of 0.02 (solar; Section \ref{sec:VLTenv}). The photometry used in the likelihood calculations was corrected for foreground galactic extinction with a value of $A_V^{MW}=0.139$ mag (NED; \citealt{2011ApJ...737..103S}) and a standard extinction law with $R_{v}=3.1$. 

\begin{figure}
\centering
\includegraphics[width=85mm]{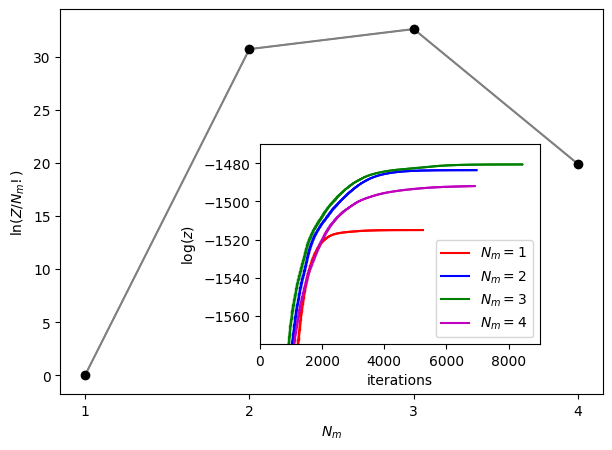}
\vspace{-2mm}
\caption{Main plot: corrected Bayesian evidence values for 1, 2, 3 and 4 age component \textsc{ultranest} runs. Subplot: evolution of evidence collection during nested sampling algorithm for each \textsc{ultranest} run.}
\label{fig:evidence}
\end{figure}

Sourcing the posterior probability distributions of our free parameters through \textsc{ultranest} requires simple function definitions for the expected form of our log-likelihoods and priors, alongside the predetermined likelihood distributions for each star, to perform the Nested Sampling algorithm. The log-likelihoods were assumed to follow a multivariate normal distribution, $N(\tau_k, A_{V,k}, \sigma_{\tau_k}, \sigma_{A_{V,k}})$, for each stellar population. Modelling two or more stellar populations requires the weight ($\omega_k$) of each distribution to be included as free parameters. For $N_m$ populations, only $N_m - 1$ weights that obey a Dirichlet distribution need to be defined in the \textsc{ultranest} log-likelihood functions. This assists in reducing the number of free parameters in our model and speeds up each run. The final weight is simply calculated using the normalisation condition $\underset{k}{\Sigma} \, {\omega}_k=1$. Flat priors are imposed on $\tau_k$, $A_{V,k}$ and $\omega_k$ with the following ranges: $6.00\leq \tau \leq8.00$, $0.00\leq A_V \leq2.00$, $0.00< \omega_k <1.00$. To further reduce the computational expense when trying to fit for a high number of population components (i.e. $N_m = 4$) initial \textsc{ultranest} runs fit for $N_m$ age components, but only a single internal extinction component and fixed age/extinction dispersions: $\sigma_{A_{V,k}}, \sigma_{\tau_k}$ = 0.05.         
With the above implementation, \textsc{ultranest} runs with 400 live points were separately performed on models with 1, 2, 3 and 4 age components. Bayesian evidence was collected and corrected for (as described in Section \ref{sec:Bayes}) following each run, the results of which are presented in Figure \ref{fig:evidence}. Also included as a subplot is the evolution of the evidence accumulation with increasing iterations, showing that the evidence converges towards a limit in each case. Following the interpretation by \cite{1939thpr.book.....J}, the largest relative jump in evidence is seen between the $N_{m}=1$ and $N_{m}=2$ age component models. Therefore, we can conclude that the photometry from the environment of iPTF13bvn is best-fit by two populations of stars representing distinct epochs of star formation. 

Following this, a secondary \textsc{ultranest} run was performed with unique extinction estimations for each population. Log-likelihood and prior functions follow the same initialisation as described above, except we almost double the number of free parameters in this comprehensive model: $\tau_1$, $\tau_2$, $A_{V,1}$, $A_{V,2}$, $\sigma_{A_{V,1}}$, $\sigma_{A_{V,2}}$. Prior ranges for each free parameter were informed by the results of the original two age component model: $6.20\leq\tau_{k}\leq7.20$, $0.40\leq A_{V,k}\leq1.20$, $0.00\leq \delta A_{V,k}\leq1.00$ (corresponding to $0.05\leq\sigma_{A_{V,k}}\leq0.50$). The extinction was chosen to have a unconstrained dispersion since this is a physical parameter that we could expect to see variation between populations, as well as a spread within each population. Inversely, we expect population ages to be tightly constrained; therefore, the age dispersion remains fixed, but is slightly relaxed to a value of $\sigma_{\tau_k} = 0.10$. 

\begin{figure*}
\centering
\includegraphics[width=175mm]{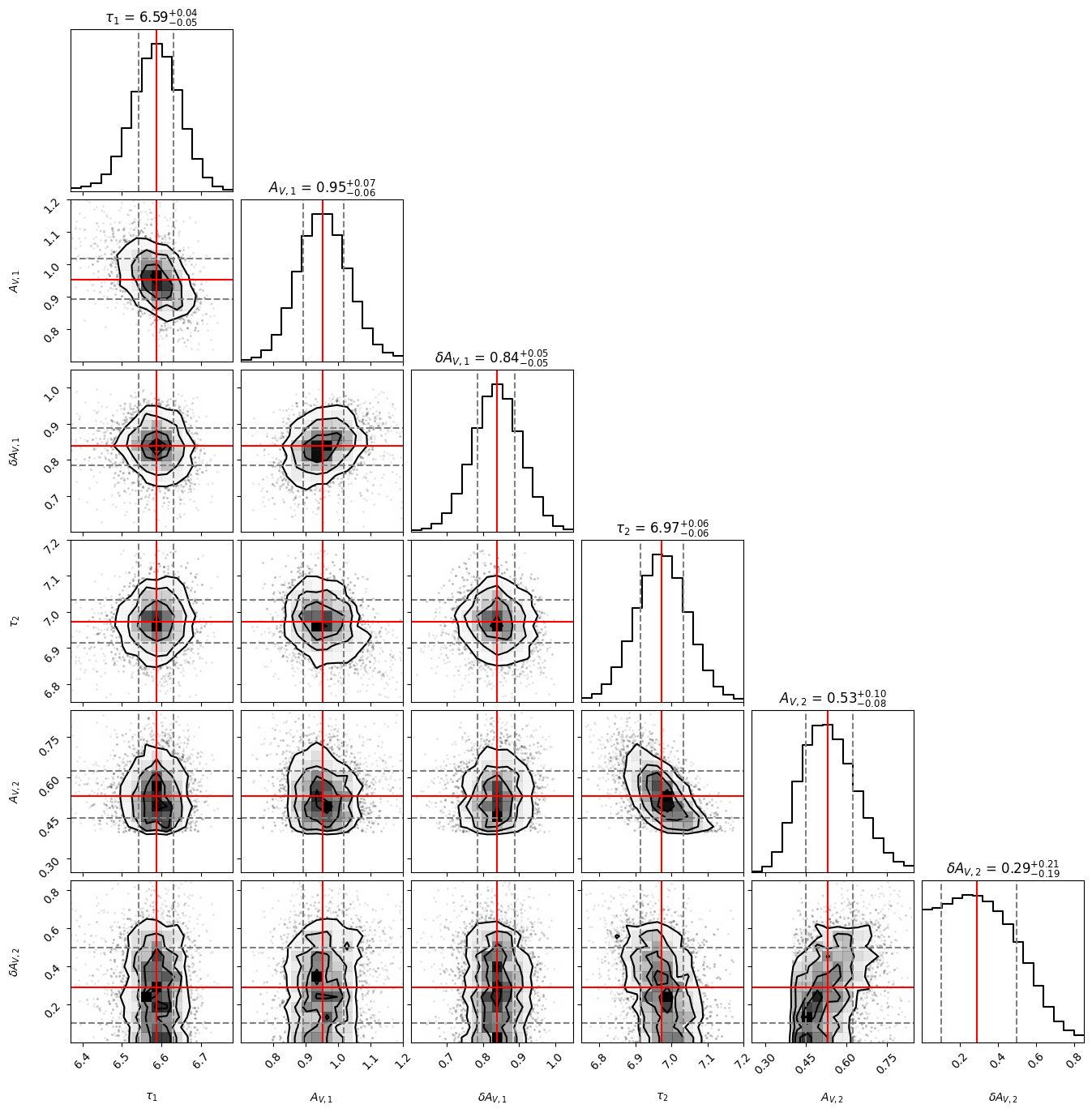}
\vspace{-2mm}
\caption{\textsc{ultranest} results using two age, $\tau_k = \log(\textrm{age}_k)$, and extinction, $A_{V,k}$, components, with a fixed age dispersion of $\sigma_{\tau_k}=0.10$. The extinction dispersion parameter $\delta A_{V,k}$, where $\sigma_{A_{V,k}}=0.05\times10^{\delta A_{V,k}}$, is left as a free parameter to be fit. Posterior distributions are plotted as histograms with the corresponding contour plots. Red lines show the best-fit values and grey dashed lines represent the 16\% and 84\% quantiles.}
\label{fig:2comp_custom}
\end{figure*}

\begin{figure*}
\centering
\includegraphics[width=175mm]{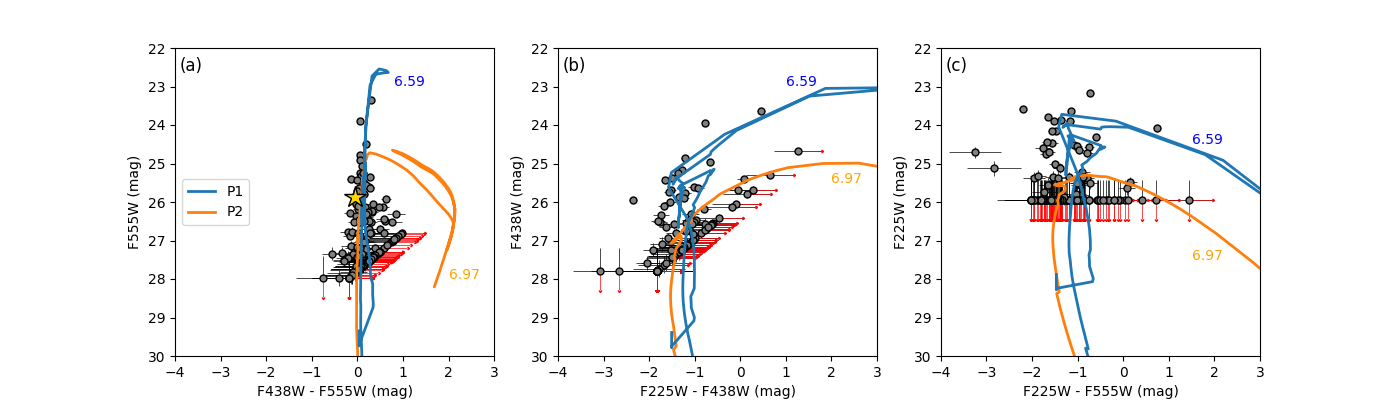}
\vspace{-2mm}
\caption{Colour magnitude diagrams (CMDs) of the 138 stars (black data points) in the environmnet of iPTF13bvn, using three colours from the HST filters: F555W, F438W, F225W. Best-fit \textsc{parsec} isochrones from the results of the two age/extinction component \textsc{ultranest} run are plotted over the data; population 1 (pop1; blue) uses $\tau_{1}=6.59$, $A_{V,1}=0.95$ mag; population 2 (pop2; orange) uses $\tau_{2}=6.97$, $A_{V,2}=0.53$ mag. Both isochrones account for foreground galactic extinction; $A_{V}^{MW}=0.139$ mag. The gold star in plot (a) shows the apparent magnitudes of the iPTF13bvn progenitor, found by \citet{2015MNRAS.446.2689E}.}
\label{fig:CMD}
\end{figure*}

Figure \ref{fig:2comp_custom} shows the results from this model run, which has successfully resolved two stellar populations with significantly different extinctions. Best-fit values for the age, extinction and corresponding dispersion are $\tau_{1}=6.59^{+0.04}_{-0.05}$, $A_{V,1}=0.95^{+0.07}_{-0.06}$ mag, $\sigma_{A_{V,1}}=0.346$ and $\tau_{2}=6.97^{+0.06}_{-0.06}$, $A_{V,2}=0.53^{+0.10}_{-0.08}$ mag, $\sigma_{A_{V,2}}=0.097$ for population 1 (P1) and population 2 (P2), respectively. Doubling the age dispersion resulted in differing peak posterior values for the population ages between our original simplified model and those seen in Figure \ref{fig:2comp_custom}. However, the P1 ages agree within $2\sigma$ and the P2 ages within $1\sigma$; given the small errors of our age estimates, such a minimal shift in values implies that relaxing the age dispersion has negligible effect on the final result. Our two population ages correspond to $\sim3.89^{+0.38}_{-0.42}$ million years for P1 and $\sim9.33^{+1.38}_{-1.20}$ million years for P2, temporally separating the epochs of star formation by $\sim5.44$ million years. The respective \textsc{parsec} isochrones for P1 and P2 are displayed in Figure \ref{fig:CMD}, overlaid on the photometry of the resolved stars (grey data points) within the r = 300 pc environment. Taking the maximum mass of the corresponding age \textsc{parsec} isochrone to be our best estimate of the progenitor mass for single star evolution, we obtain the following initial (zero-age main sequence) and final masses:  
\begin{itemize}\centering
\item[] $M_\textrm{initial,P1} = 65.4 \textrm{ M}_\odot$.
\item[] $M_\textrm{final,P1} = 20.1 \textrm{ M}_\odot$.
\item[] 
\item[] $M_\textrm{initial,P2} = 20.0 \textrm{ M}_\odot$.
\item[] $M_\textrm{final,P2} = 17.8 \textrm{ M}_\odot$.
\end{itemize}
\textsc{parsec} final masses reflect mass-loss across advanced evolutionary phases, as late as central carbon ignition \citep{2015MNRAS.452.1068C}. Since this does \textit{not} model rapid mass-loss in the final centuries of massive stars, nor mass-loss through binary interaction, the final masses do \textit{not} represent pre-explosion masses. Designating a population age and mass to the progenitor of iPTF13bvn is subject to interpretation and is discussed further in Section \ref{sec:HostPop}. In short, although we cannot conclusively assign a natal stellar population, the collective environmental information for iPTF13bvn motivates a statistical comparison to other Type Ib CCSNe, as well as other classifications in a future sample study. 

\subsection{The Environment of iPTF13bvn as seen through VLT/MUSE}
\label{sec:VLTenv}
Running the algorithm described in Section \ref{sec:algorithm}, fitting the continuum and H$\alpha$/$\beta$ Balmer lines of individual spaxels in our $30\times30$ datacube, produced a myriad of parameter maps displayed in Figure \ref{fig:muse}.  From the top row of plots (mean continuum; H$\alpha$ flux; H$\beta$ flux) we can see the distribution of white-light sources against numerous bright H \textsc{ii} regions. However, the immediate site of iPTF13bvn is comparatively dark despite being coincident with the spiral arm; instead it is encircled by H \textsc{ii} regions. Where we suspect there is a dust lane arching around the site of iPTF13bvn from a dark feature in Figure \ref{fig:environment} coincides with a drop in signal in the MUSE data, specifically where we see missing spaxels inside the spiral arm (Figure \ref{fig:muse} [c]). Looking at rows 2 and 3 of the first column of Figure \ref{fig:muse}, our ‘best fitting’ numeric (outlined in Section \ref{sec:algorithm}) shows contrasting results for both Balmer lines. Fitting of H$\alpha$ favours the ‘weak’ single Gaussian (3) function, with only the brightest H \textsc{ii} regions optimised by the ‘strong’ single Gaussian (4). Comparatively, fitting of H$\beta$ applies the full range of $(1) \rightarrow (6)$ fitting functions, producing identifiable regions and borders within the datacube environment. Mainly, the spiral arm becomes visible and is overwhelmingly fit with either ‘weak’ or ‘strong’ double Gaussian functions (1 \& 2), whilst some spaxels on the arm-edge are enforced with ‘weak’ single Gaussian fittings (3) due to decreasing SNR. Whilst the dual absorption-emission features across the spiral arm are correlated with cospatial stars and gas, the ‘pure absorption’ (5) fitting seen in the top left corner indicates dense (likely older) stellar populations external to the spiral arm. The remaining spaxels are mostly fit with a featureless ‘continuum’ (6). 

Given the observed correlation between EW(H$\alpha$) and the specific star-formation rate (sSFR) \citep{2016MNRAS.460.3587M}, we see that the brightest H \textsc{ii} regions correspond to the highest sSFR. Overall, we can summarise the environment of iPTF13bvn to be: within a spiral arm of NGC 5806; lacking in ongoing star-formation; comparatively deficient in hard radiation from massive star populations.  

Additionally, a gas-phase metallicity estimate for the environment of iPTF13bvn (‘13bvn-outer’ in Figure \ref{fig:muse}) was found using a strong-line diagnostic, specifically the O3N2 calibration by \citet{2013A&A...559A.114M}. From the summed spectrum within this aperture, emission line fluxes for the [O III] $\lambda5007$ and [N II] $\lambda6584$ lines were estimated from a single Gaussian fit (H$\alpha$ and H$\beta$ fluxes are estimated as standard by the fitting algorithm). Through the [O III] $\lambda5007$/H$\beta$ and [N II] $\lambda6584$/H$\alpha$ line ratios, we find an abundance estimate of $12 + \log{(\textrm{O/H})} = 8.69 \pm 0.18$ dex. This matches the current solar value (8.69 dex; \citealt{2009ARA&A..47..481A}), therefore all prior assumptions of a solar metallicity in our analyses are valid.  

\subsection{Extinctions from MUSE Spectral Analysis}
\label{sec:MUSEresults}

\begin{figure*}
\centering
\includegraphics[width=175mm]{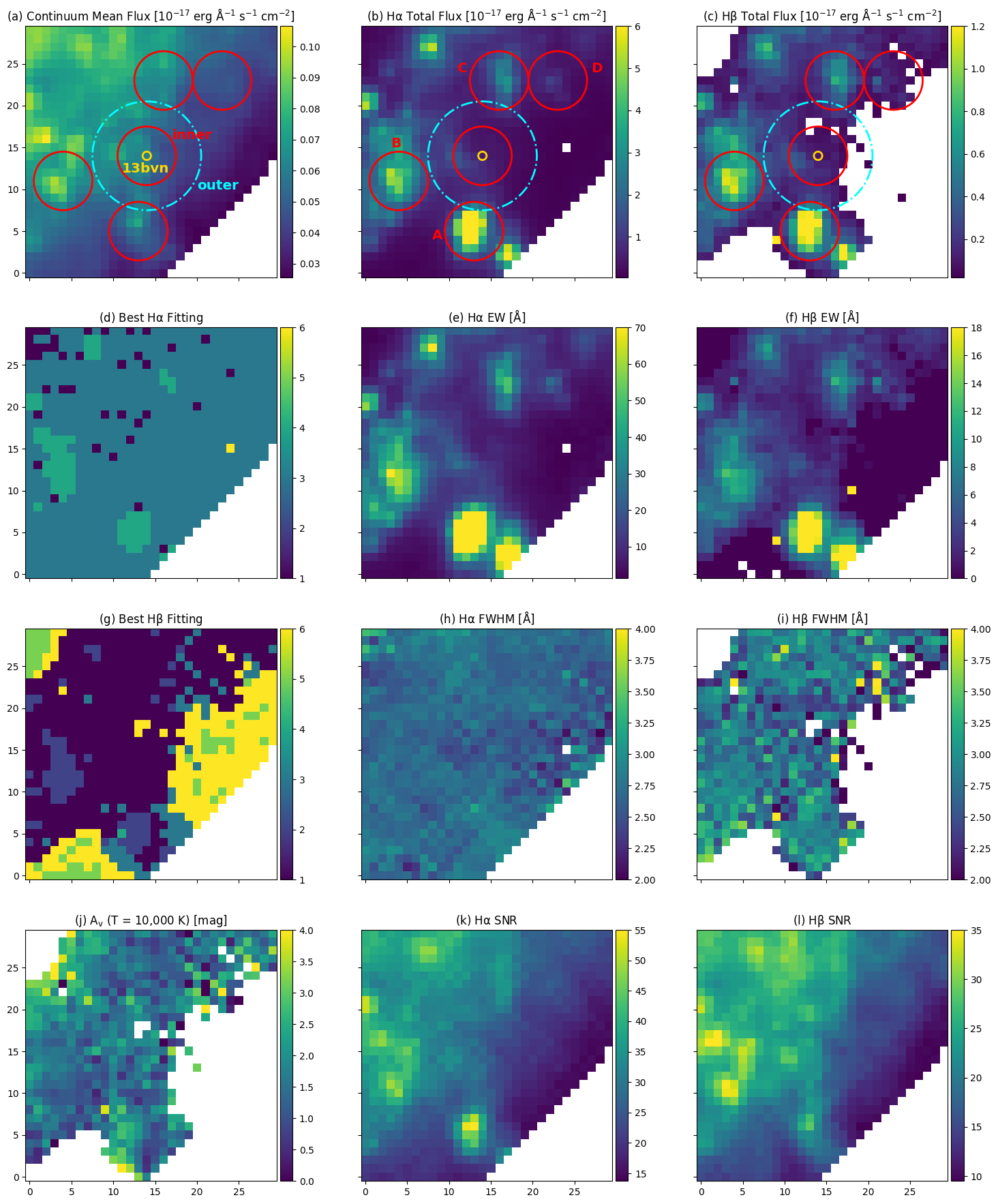}
\vspace{-2mm}
\caption{Maps of derived spectral observables for our extracted $30\times30$ rebinned spaxel MUSE datacube (Figure \ref{fig:environment}). These include: (a) mean continuum flux; (b)/(c) emission line flux of H$\alpha$/$\beta$ respectively; (d)/(g) 'best fitting' numeric (Section \ref{sec:algorithm}) for H$\alpha$/$\beta$ respectively; (e)/(f) emission line equivalent width (EW) for H$\alpha$/$\beta$ respectively; (h)/(i) emission line full width at half maximum (FWHM) for H$\alpha$/$\beta$ respectively; (j) Balmer decrement extinctions ($A_V$) with $I[\textrm{H}\alpha]/I[\textrm{H}\beta] = 2.86$, T = 10,000 K and $R_V = 3.1$; (k)/(l) emission line signal-to-noise ratio (SNR) for H$\alpha$/$\beta$ respectively. Overlaid are circular apertures for '13bvn\_inner' (radius=3.5 spaxels) and '13bvn\_outer' (radius=6.5 spaxels), as well as comparison environments (A, B, C and D) with radii=3.5 spaxels (refer to Table \ref{tab:extinctions}). The site of iPTF13bvn lies at the centre of the image (14,14) and is labelled in gold.} 
\label{fig:muse}
\end{figure*}

\begin{figure}
\centering
\includegraphics[width=85mm]{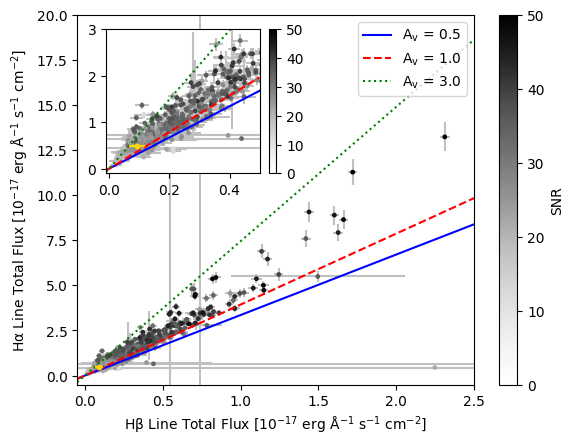}
\vspace{-2mm}
\caption{Main plot: Balmer line (H$\alpha$ and H$\beta$) emission fluxes of individual rebinned spaxels from our extracted MUSE datacube, and corresponding error estimates. Greyscale colourbar on datapoints represents the H$\alpha$ line SNR. Spaxel (14,14) in Figures \ref{fig:muse} and \ref{fig:hst_resampled}, containing the location of iPTF13bvn, is highlighted as a golden datapoint. Subplot: zoom-in of bottom left corner of main plot.} 
\label{fig:extinctions}
\end{figure}

Implementation of the Balmer decrement method to acquire extinction estimates across the environment of iPTF13bvn was applied to individual spaxels, as well as the summed spectra of spaxels contained within a defined region. Since the electron density and temperature are not constrained in this study, any single extinction estimate comprises a set of four values to cover the full temperature range (2500 - 20000 K) with independent intrinsic $I[\textrm{H}\alpha]/I[\textrm{H}\beta]$ ratios from \citet{2006agna.book.....O}. Figure \ref{fig:extinctions} shows the distribution of single spaxel extinctions (T = 10,000 K) throughout the datacube, where the majority of spaxels are constrained between $A_V = 0.5$ mag and $A_V = 3.0$ mag. Propagating our flux error estimates produces reasonable extinction errors for most of this distribution, even in the low ($< 15$) SNR regime. The few exceptions to this, with large errors, highlight the limitations of our fitting algorithm where low H$\beta$ flux results in poorly constrained Gaussian fits. Further confidence in the performance of our fitting algorithm can be gained by the smoothly varying nature of the spatial map of extinctions in Figure \ref{fig:muse}(j). For the single spaxel containing the position of iPTF13bvn, the extinction estimate (assuming $T = 10000\, \mathrm{K}$) is shown as a golden datapoint in Figure \ref{fig:extinctions} and corresponds to a value of:
\begin{itemize}\centering
\item[] $A_{V, \textrm{13bvn-single}} = 1.92 \pm 0.13$ mag. 
\end{itemize}
Given our pair of population results from the photometric-isochrone fitting, an equivalent environment definition (circular aperture; $r = 300\,\mathrm{pc}$; centred on iPTF13bvn) was applied to the MUSE datacube. All spaxels which fall inside this aperture had their spectra summed and the fitting algorithm was applied to the product. The resulting extinction estimate can be interpreted as a single averaged value across the environment (env):
\begin{itemize}\centering
\item[] $A_{V, \textrm{13bvn-env}} = 1.74 \pm 0.05$ mag. 
\end{itemize}
To test the reliability of our single spaxel extinctions we find a weighted mean ($\bar{A}_{V}$) for the sample within the r = 300 pc environment, with the expectation that this should reproduce the summed environmental value. This allows us to gain a new parameter, the standard error ($\sigma_{\bar{A}_{V}}$), representing the dispersion about this weighted mean. For our photometric-equivalent (r = 300 pc) environment, this produces:
\begin{itemize}\centering
\item[] $\bar{A}_{V, \textrm{13bvn}} = 1.463$ mag.
\item[] $\sigma_{\bar{A}_{V, \textrm{13bvn}}} = 0.005$ mag. 
\end{itemize}
The above set of extinctions are significantly different, the only agreement can be found between $A_{V, \textrm{13bvn-single}}$ and $A_{V, \textrm{13bvn-env}}$ to within $2\sigma$. The individual spaxel distributional mean ($\bar{A}_{V, \textrm{13bvn}}$) does \textit{not} reproduce the stacked spectrum estimate ($A_{V, \textrm{13bvn-env}}$) across the $r = 300$ pc environment of iPTF13bvn. We would \textit{not} expect this disagreement between $\bar{A}_{V, \textrm{13bvn}}$ and $A_{V, \textrm{13bvn-env}}$ unless there was a sharp drop off in flux within the spaxels that comprise the $r=300$ pc aperture. This prompts an investigation into whether this result is systematically observed across the MUSE datacube or unique. Although varying the electron temperature shifts these extinctions values (discussed further in Section \ref{sec:temp_dependence}), the disagreement between these two values remains.

\subsection{Comparison Environments}
\label{sec:comparison_envs}

The ensemble of extinction estimates for the environment of iPTF13bvn requires context within the larger environment around iPTF13bvn. Selected comparison environments are shown in Figure \ref{fig:muse}, where regions A, B and C are the nearest H \textsc{ii} regions (A \& B are considerably brighter than C) and region D was selected for its likeness to the environment of iPTF13bvn. These regions were selected based on the ionsed gas H$\alpha$ and H$\beta$ flux, centred on the brightest spaxels of the MUSE data. An additional region within the environment of iPTF13bvn was created, ‘13bvn-inner’, that provides an equivalent aperture to the comparison environments (A, B, C and D). Our photometric-equivalent environment is now labelled as ‘13bvn-outer’. Region D and ‘13bvn-inner’ both lie on the edge of the spiral arm, with relatively low flux levels in continuum and emission. The smaller of the apertures have radii of 3.5 rebinned MUSE spaxels ($\sim 150$ pc), whereas the ‘13bvn-outer’ aperture has r = 6.5 rebinned MUSE spaxels ($\sim 300$ pc). 

\begin{table*}
\caption{Summary of extinction results for all environments (iPTF13bvn and comparisons) defined in Figure \ref{fig:muse} including: the single extinction estimate of the summed spectrum for each region ($A_{V, \textrm{env}}$), the weighted-mean of the distribution of single spaxel extinction estimates within each region ($\bar{A}_{V}$) and the corresponding standard error of each weighted-mean ($\sigma_{\bar{A}_{V}}$). Projected distances to iPTF13bvn are found from the SN site spaxel (14,14) to the central spaxel of each region, it is worth noting that regions A through D have aperture radii of $\sim 152.6$ pc (3.5 MUSE spaxels). The final column shows the EW equivalent $\log(age\, \textrm{years})$ for \textsc{starburst99} models (instantaneous; $Z=0.02$); errors are not included as our EW errors are outside the resolution of the models.} 
\label{tab:extinctions}
{\begin{tabular}{c|c|c|c|c|c|c|c}
    \hline
    Region & \vtop{\hbox{\strut Aperture Radius}\hbox{\strut [pc] (spaxels)}} & \vtop{\hbox{\strut $A_{V,\textrm{env}}$ [mag]}\hbox{\strut (T=10kK)}} & $\bar{A}_{V}$ [mag] & $\sigma_{\bar{A}_{V}}$ [mag] & \vtop{\hbox{\strut Projected Distance}\hbox{\strut to iPTF13bvn [pc]}} & H$\alpha$ EW [\AA] & \vtop{\hbox{\strut \textsc{starburst99}}\hbox{\strut $\log(age\, \textrm{years})$}}\\
    \hline\hline
    iPTF13bvn outer & 283 (6.5) & $1.738 \pm 0.052$ & 1.467 & 0.005 & 0.0 & $13.5 \pm 1.5$ & 7.04 \\ 
    \hline
    iPTF13bvn inner & 153 (3.5) & $1.479 \pm 0.060$ & 1.423 & 0.011 & 0.0 & $12.5 \pm 1.5$ & 7.05 \\
    \hline
    A & 153 (3.5) & $1.832 \pm 0.109$ & 1.760 & 0.005 & 394.8 & $78.6 \pm 19.0$ & 6.80 \\
    \hline
    B & 153 (3.5) & $1.399 \pm 0.061$ & 1.384 & 0.004 & 455.2 & $41.5 \pm 5.9$ & 6.81 \\
    \hline
    C & 153 (3.5) & $1.361 \pm 0.047$ & 1.407 & 0.006 & 402.0 & $23.4 \pm 2.4$ & 6.86 \\ 
    \hline
    D & 153 (3.5) & $1.770 \pm 0.065$ & 1.775 & 0.016 & 554.9 & $11.3 \pm 1.4$ & 7.05 \\  
    \hline
\end{tabular}}
\end{table*}

Values for $A_{V, \textrm{env}}$, $\bar{A}_{V}$ and $\sigma_{\bar{A}_{V}}$ were determined for every defined region in Figure \ref{fig:muse}, the results of which are summarised in Table \ref{tab:extinctions}. Notably, all regions besides ‘13bvn-outer’ find good agreement, within $1\sigma$, between their environmentally stacked spectrum estimate and their single spaxel distributional mean estimate. Following the question raised in Section \ref{sec:MUSEresults} this result for '13bvn-outer' is unique to the comparison environments, implying that the larger aperture (increased number of spaxels) of '13bvn-outer' results in the disagreement between $A_V$ and $\bar{A}_{V}$. Furthermore, the weighted mean standard error, $\sigma_{\bar{A}_{V}}$, for ‘13bvn-inner’ and region D are twice what is found for the remaining environments. 

\subsection{H \textsc{ii} Region Ages with \textsc{starburst99}}
\label{sec:starburst99}

Following the standard practice to estimate the ages of our bright H \textsc{ii} regions using the equivalent widths (EWs) of their H$\alpha$ profiles, we directly compare our measurements to the \textsc{starburst99} 1999 tabular dataset \citep{1999ApJS..123....3L}. The \textsc{starburst99} models synthesize instantaneous star formation (i.e. a singular stellar population) and implement stellar evolution models by the Geneva group \citep{1990A&AS...84..139M, 1992A&AS...96..269S, 1993A&AS..102..339S, 1993A&AS...98..523S, 1993A&AS..101..415C} across a range of discrete metallicities ($Z = 0.040, 0.020, 0.008, 0.004\, \textrm{and}\, 0.001$). The EWs were derived from the summed spectra of the defined apertures in Section \ref{sec:comparison_envs}, for the three closest bright H \textsc{ii} regions to the site of iPTF13bvn: Regions A, B and C. These were compared to the instantaneous SF \textsc{starburst99} model with $Z = 0.02$ (their analogue to $Z_\odot$), a \citet{1955ApJ...121..161S} IMF with a power law exponent $\alpha = 2.35$ and steps of 100,000 years. Our EW values are listed in Table \ref{tab:extinctions}, along with the \textsc{starburst99} model ages for the closest matching EW. Since they consider $Z_\odot = 0.02$, which is higher than current day best-estimates ($Z_\odot = 0.0134$; \citealt{2009ARA&A..47..481A}), we also compare to the $Z = 0.008$ models, but the difference in resulting age estimates is negligible for the purposes of this comparison. The corresponding $\tau_\textrm{env}=\log(age\, \textrm{years})$ are $\tau_\textrm{A}=6.80$, $\tau_\textrm{B}=6.81$ and $\tau_\textrm{C}=6.86$; here each $\tau_\textrm{env}$ value represents a unique age estimate from independent H \textsc{ii} regions (refer to Figure \ref{fig:muse}). Errors are not stated, as our EW errors land inside the age resolution of the \textsc{starburst99} models. All three ages fall between our photometrically derived stellar population values ($\tau_\textrm{P1}$ and $\tau_\textrm{P2}$; Section \ref{sec:bayesresults}).          

\subsection{Additional Information from Resampled HST-to-MUSE Data} 
\label{sec:resampleresults}

\begin{figure*}
\centering
\includegraphics[width=175mm]{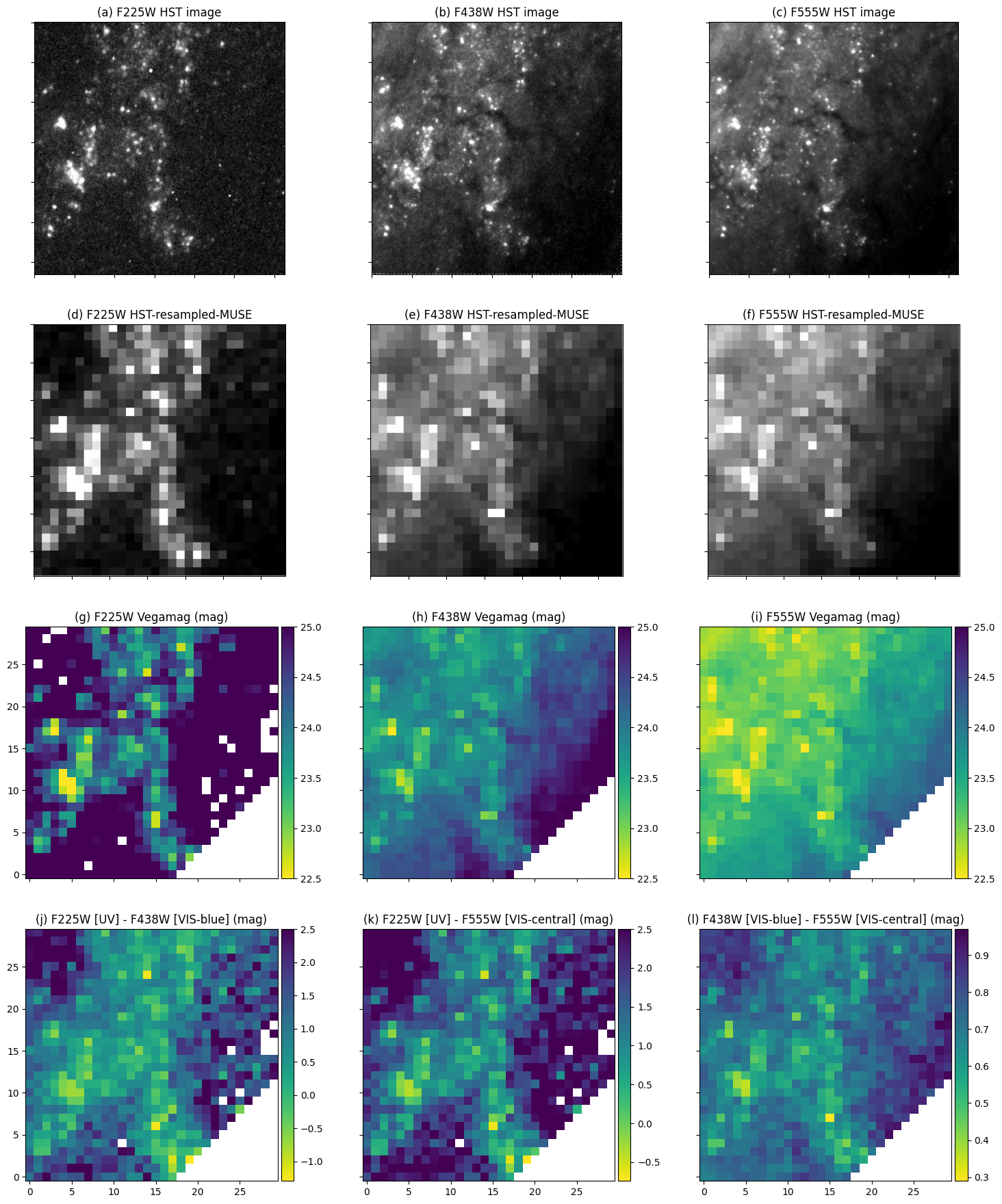}
\vspace{-2mm}
\caption{Row by row stages of transforming HST images into Vega magnitude and colour-colour maps with the same spatial sampling as our $30\times30$ rebinned spaxel MUSE datacube. Columns correspond to a given filter: F225W (left), F438W (central) and F555W (right). First row: HST (\_drc.fits) images. Second row: raw spatially-resampled images. Third row: electron counts have been converted into Vega magnitudes (Section \ref{sec:resample}). Fourth row: colour-colour maps.} 
\label{fig:hst_resampled}
\end{figure*}

Given the advantageous 0.04 arcsec pixel scale of the HST/WFC3 data, we can see how resolved stars and clusters transform onto the larger (0.2 arcsec) pixel map of the VLT/MUSE datacube. Conversion of electron counts into surface brightness allows us to visualise these resampled images as magnitude and colour maps, presented in Figure \ref{fig:hst_resampled}. The ability of our fitting numeric (Figure \ref{fig:muse}; ‘Best H$\beta$ Fitting’) to outline the galactic spiral arm is traced, almost one-to-one, by the brightest UV (F225W) sources. Despite being brightest in pixels cospatial with our brightest H$\alpha$ flux (H \textsc{ii} regions in Figure \ref{fig:muse}), the UV flux is observed across the entire spiral arm. This F225W Vega magnitude map also reconfirms the harsh edge in ionising sources close to the site of iPTF13bvn. 

\section{DISCUSSION}
\label{sec:discussion}

\subsection{Host Resolved Population (HST)}
\label{sec:HostPop}

\begin{table*}
\caption{Summary of results from past studies of iPTF13bvn.} 
\label{tab:13bvn}
\resizebox{\textwidth}{!}{\begin{tabular}{l|l|l|l|l|l}
    \hline
    Reference & Distance (Mpc) & Age ($\log{(t\, \textrm{years})}$) & Local Extinction & Progenitor & Methods \\ [0.5ex]
    \hline\hline
    \cite{2013ApJ...775L...7C} & 22.5 & - & \vtop{\hbox{$E(B-V)=0.0437$}\hbox{($A_V \approx 0.14$ mag $\dagger$)}} & Single; Wolf-Rayet & \vtop{\hbox{\strut Extinction: EW of Na I D line.}\hbox{\strut Progenitor: pre-explosion detection limits,}\hbox{\strut mass-loss estimate from radio data.}} \\ 
    \hline
    \cite{2014AJ....148...68B} & 25.54 & - & \vtop{\hbox{$E(B-V)=0.17$}\hbox{($A_V \approx 0.53$ mag $\dagger$)}} & \vtop{\hbox{\strut Binary; $\sim3.5\, \textrm{M}_\odot$ pre-explosion}\hbox{\strut low-mass helium star;}\hbox{\strut initial mass $\sim20\, \textrm{M}_\odot$.}} & \vtop{\hbox{\strut Extinction: comparing corrected $E(B-V)$}\hbox{\strut colours to an intrinsic colour law derived}\hbox{\strut from a sample of SESNe.}\hbox{\strut Progenitor: hydrodynamical modelling}\hbox{\strut of lightcurve.}} \\
    \hline
    \cite{2015MNRAS.446.2689E} & 22.5 & - & $E(B-V)=0.0437 - 0.17$ & \vtop{\hbox{\strut Binary; low-mass helium star;}\hbox{\strut initial mass $\sim10-20\, \textrm{M}_\odot$.}} & \vtop{\hbox{\strut Extinction: previous studies.}\hbox{\strut Progenitor: revised photometry,}\hbox{\strut SED model fitting.}} \\
    \hline
    \citet{2015AA...579A..95K} & - & - & - & \vtop{\hbox{\strut Binary; oxygen mass $\leq0.7 M_\odot$,}\hbox{\strut initial mass $\leq15-17 M_\odot$}} & \vtop{\hbox{\strut Progenitor: spectroscopic emission line}\hbox{\strut fluxes and ratios.}} \\
    \hline
    \cite{2016MNRAS.461L.117E} & 22.5-26.8 & - & $E(B-V)=0.0437 - 0.17$ & \vtop{\hbox{\strut Binary; initial mass $10-12 M_\odot$}\hbox{\strut helium giant.}} & \vtop{\hbox{\strut Extinction: previous studies.}\hbox{\strut Progenitor: SED fitting to BPASS models.}} \\ 
    \hline
    \cite{2016ApJ...825L..22F} & 25.8 & - & $E(B-V)=0.17$ & Binary; non-definitive. & \vtop{\hbox{\strut Extinction: previous studies.}\hbox{\strut Progenitor: obtained deep HST photometry}\hbox{\strut in nebular phase, fit SED to models.}} \\  
    \hline
    \cite{2018MNRAS.476.2629M} & 26.10 &  \vtop{\hbox{\strut$\tau_{1}=6.74\pm0.02$} \hbox{\strut$\tau_{2}=7.38^{+0.01}_{-0.06}$}} & $A_V=0.98\pm0.02$ & Initial mass $10.5^{+1.0}_{-0.7} - 36^{+9}_{-6}\, \textrm{M}_\odot$ & \vtop{\hbox{\strut Extinction and Age: hierarchical Bayesian}\hbox{\strut mixture model fitting.}} \\ 
    \hline
    \cite{2023MNRAS.521.2860S} & 27.0 & $\tau=6.71\pm0.04$ & $A_V=0.78\pm0.13$ & Initial mass $39.9^{+6.9}_{-5.2}\, \textrm{M}_\odot$ & \vtop{\hbox{\strut Extinction and Age: hierarchical Bayesian}\hbox{\strut mixture model fitting.}} \\
    \hline
    \textbf{This Work} & 22.5 &  \vtop{\hbox{\strut$\tau_{1}=6.59^{+0.04}_{-0.05}$} \hbox{\strut$\tau_{2}=6.97^{+0.06}_{-0.06}$}} & \vtop{\hbox{\strut$A_{V,1}=0.95^{+0.07}_{-0.06}$} \hbox{\strut$A_{V,2}=0.53^{+0.10}_{-0.08}$}} & Initial mass $20.0 - 65.4\, \textrm{M}_\odot$ & \vtop{\hbox{\strut Extinction and Age: hierarchical Bayesian}\hbox{\strut mixture model fitting.}} \\
    \hline 
\multicolumn{6}{l}{$\dagger$ Conversion from $E(B-V)$ to $A_V$ assumes an $R_V=3.1$.}\\ 
\end{tabular}
}
\end{table*}

Identifying the host population of iPTF13bvn from our resolved stellar populations (P1 \& P2) remains inconclusive. \cite{2023MNRAS.521.2860S} used the same approach to population fitting, but in the cases where more than one population were recovered the youngest population was preferentially selected, as these populations represent the greatest contributors to the UV flux at the centre of their study. Since our data extends across a larger wavelength coverage, the same assumption will not be made. The probabilities of each star belonging to either resolved stellar population in the environment of iPTF13bvn were found in an attempt to spatially distinguish the two populations. As was found by \citet[][; see their Fig. 10]{2018MNRAS.476.2629M} for SN 1993J, we find an even blend of stars belonging to both populations across the environment of iPTF13bvn, therefore it is \textit{not} viable to spatially separate either population as an approach to identifying a host population. A growing consensus in binary interaction being the predominant mass-loss channel for stripped envelope SESNe \citep{2014ARA&A..52..487S, 2019sros.confE..73S} may advocate for the older population to be the stronger choice; binary mass exchange allows lower mass (longer lived) progenitors to successfully explode and the IMF favours lower mass progenitors. 

In the case of iPTF13bvn, we are afforded the luxury of a reasonably well constrained progenitor from over a decade of studies. The results of which are condensed into Table \ref{tab:13bvn} and can assist in constraining a host population. \citet{2018MNRAS.476.2629M} and \citet{2023MNRAS.521.2860S} are past results that employed the same Bayesian methodology as in this paper. A key difference is that \citet{2018MNRAS.476.2629M} used three HST/ACS/WFC1 filter (F435W; F555W; F814W) images, with comparatively short exposure times and a wavelength coverage mostly contained to the visible. \citet{2023MNRAS.521.2860S} focused their study on the UV data of SESNe with two HST/WFC3/UVIS filter (F300X and F475X) images for iPTF13bvn. Our wavelength range encompasses those of both studies and our images achieve deeper detection limits. Of the two stellar populations recovered by \citet{2018MNRAS.476.2629M}, the youngest most closely resembles our P1 result, particularly their estimate for the extinction ($A_V = 0.98 \pm 0.02$ mag). The single young population found by \citet{2023MNRAS.521.2860S} agrees with our P1 age estimate (within $3\sigma$). Their extinction ($A_V = 0.78 \pm 0.13$ mag) estimate falls in-between P1 and P2, but their relatively large error means it is within $2\sigma$ of both values. This supports the need to prescribe two distinct stellar populations to the environment of iPTF13bvn in order to produce well constrained extinction estimates. 

Excluding \citet{2018MNRAS.476.2629M} and \citet{2023MNRAS.521.2860S}, the remaining studies rely on two extinction estimates: \citet{2013ApJ...775L...7C} $E(B-V) = 0.0437$ ($A_V \approx 0.14$ mag) and \citet{2014AJ....148...68B} $E(B-V) = 0.17$ ($A_V \approx 0.53$ mag). Furthermore, the progenitor converges on a binary system with a primary initial mass within the range of $10 - 20\, \textrm{M}_\odot$, ending its life as a low-mass helium star. This aligns with our results for P2: $A_{V,P2} = 0.53^{+0.10}_{-0.08}$ mag; $M_{initial,P2} = 20.0\, \textrm{M}_\odot$ (zero-age main sequence); $M_{final,P2} = 17.8\, \textrm{M}_\odot$ (post advanced evolutionary stages; Section \ref{sec:bayesresults}). \citet{2018MNRAS.476.2629M} finds an older population with $\tau = 7.38^{+0.01}_{-0.06}$, which does not agree with our P2 result. Although their corresponding progenitor mass estimate ($10^{+1.0}_{-0.7}\, \textrm{M}_\odot$) still falls within the $10 - 20\, \textrm{M}_\odot$ predicted mass for iPTF13bvn. It is worth noting that the \textsc{parsec} stellar evolution tracks are computed for single stars. Although the final mass estimates incorporate mass-loss throughout the main sequence and final evolutionary stages, mass lost to binary interaction is neglected. Including binary stripping of the primary star would result in a much lower final mass, as is concluded by \citet{2014AJ....148...68B} through hydrodynamical modeling of the lightcurve of iPTF13bvn (best fit by a $\approx 3.5\, \textrm{M}_\odot$ helium star). Our P2 extinction also agrees with the \citet{2014AJ....148...68B} value, which they estimated by comparing their $(B-V)$ colours (corrected for Milky Way reddenning) to an intrinsic colour law derived from a sample of SESNe observed by the Carnegie Supernova Project. Overall, the results of this paper are able to reproduce the populations found by \citet{2018MNRAS.476.2629M} and \citet{2023MNRAS.521.2860S}, as well as a new population that agrees with the progenitor constraints for iPTF13bvn from the remaining studies in Table \ref{tab:13bvn}. Despite this implying that P2 should be considered the host population for iPTF13bvn, we still do not have an independent check for applications to other CCSNe environments in the future. 

Plotting a CMD of the objects and isochrones corresponding to the resolved populations in the environment of iPTF13bvn, shown in Figure \ref{fig:CMD}, allows us to see which stars contribute to each fitting. This can highlight the strength of each population fitting. For P2 in plots (b) and (c) of Figure \ref{fig:CMD}, only a handful of data points with complete photometry (in all filters) constrain this older population in the region where the isochrones part. The majority of sources that assist the fitting here are detection limit substitutions. It makes sense that the older (cooler) stars appearing in the F555W and F438W images are missing from F225W, but it must be acknowledged that they are weaker probes of this portion of the CMD. It also follows that \citet{2023MNRAS.521.2860S} does not find an older population when solely using UV images. Besides the differences in our studies discussed above, \citet{2018MNRAS.476.2629M} also found detection limits for individual stars to account for variations in background and crowding. This difference in method should not massively impact how our older populations are fit, instead the main driving force behind the difference in our age estimates is likely the choice of HST images and filters. Since these cooler stars always lie close to or below detection limits, the choice of band-width, central wavelength, throughput and image depth (exposure time) all greatly impact how many of these objects are detected. In this sense, our Bayesian approach is more sensitive to fitting younger populations compared to older. 

\citet{2015MNRAS.446.2689E} found apparent magnitudes of the detected iPTF13bvn progenitor candidate in HST/ACS/WFC F435W and HST/WFC3 F555W filters (VEGAMAG system; using \textsc{dolphot}): $m_{F435W} = 25.81 \pm 0.06$ mag and $m_{F555W} = 25.86 \pm 0.08$ mag. Considerable overlap between the F435W (WFC) and F438W (WFC3) band-widths and throughputs allow us to substitute the former for the latter; both magnitudes are included in plot (a) of Figure \ref{fig:CMD}. Reassuringly, it falls upon the space occupied by the brightest stars within the field, but overlaps with the isochrones of both P1 and P2.

\subsection{Sources of Variation in Extinction (MUSE)}
\label{sec:sources_of_variation}

\subsubsection{Systematic Uncertainties}
\label{sec:syst_unc}

When interpreting the extinctions produced through our analysis of the MUSE IFU spectroscopy data, it is important to consider the sources of uncertainty and any assumptions. Systematic uncertainties include:
\begin{enumerate}
    \item distance to NGC 5806;
    \item inclination;
    \item metallicity;
    \item spectral resolution;
    \item integrated flux of emission lines;
    \item temperature dependence of extinction;
    \item and choice of $R_V$. 
\end{enumerate}
As introduced in Section \ref{sec:algorithm}, the ionised gas temperature is not assumed to be a singular value across the spiral arm of NGC 5806. Instead, we consider the full \citet{2006agna.book.....O} range spanning T = 2500 - 20000 K. We can expect different extinctions to be found when studying the stars in the environment compared to the ionised gas, as both can act as probes at varying depths through the galactic plane. In this sense, agreement or disagreement between values found with these independent probes can inform the 3D structure along our line of sight (LoS). Inclination also affects the angle of incidence between our LoS and the galactic plane, increasing the cylindrical volume of the galaxy that we observe along any LoS with decreasing inclination. This effect is more dramatic for galaxies approaching edge-on. In the case of NGC 5806, its inclination of $i = 60.4$ deg negates this effect and so is considered negligible. All quoted distances within NGC 5806 stated in this paper are projected distances, not corrected for inclination. Our choice of a 22.5 Mpc distance to NGC 5806 has negligible effect on our spectroscopic analysis, as the induced wavelength shift from varying this distance value is minuscule. Distance estimates to NGC 5806 reported on NED extend to 27.4 Mpc, the effects of which are explored by \citet{2016MNRAS.461L.117E}. To summarise, increasing the distance to NGC 5806 would have the greatest impact on our resolved stellar population analysis in Section \ref{sec:bayesresults}. Increasing this distance would increase the absolute magnitudes of individual stars and our progenitor mass estimate (decrease our stellar population age). This increase in progenitor mass would be small, nevertheless it suggests our $20.0\,\textrm{M}_\odot$ estimate should be treated as a lower limit. Uncertainty in the integrated flux of emission lines produced by our fitting algorithm are minimised by the various methods described in Section \ref{sec:algorithm} to discourage overfitting any spectra. Where it is necessary (mainly in the low SNR regime) we account for stellar absorption in order to better approximate the emission flux. In the rare case that the fitting algorithm catastrophically fails, mainly when attempting to fit the lowest flux emission, we can identify them through their unphysical error estimates and reject the data accordingly. The main limitation of our derived fluxes stems from the resolution of the MUSE spectra.

\subsubsection{Temperature Dependence of Extinction Law}
\label{sec:temp_dependence}

Intuitively, we should expect there to be variation in gas temperature when considering the $\sim 1.7\, \textrm{kpc}^2$ surface area of NGC 5806 in our extracted MUSE datacube (Figure \ref{fig:muse}). The ionised gas electron temperature can be estimated with the [O I] $(\lambda6300+\lambda6363)/\lambda5577$ or [N II]$(\lambda6548+\lambda6584)/\lambda5755$ ratios. Unfortunately, neither the [O I] $\lambda5577$ nor the [N II] $\lambda5755$ lines are detected in the single spaxel or summed spectra of iPTF13bvn; indicative of low temperatures. Our environmental and single-spaxel extinction estimates for iPTF13bvn approach our resolved stellar population values with decreasing gas temperature. Even when a 2500-5000 K temperature range is assumed, these values still significantly differ within their respective errors and the greatest discrepancy corresponds to the single-spaxel of iPTF13bvn. Although there are various interpretations (Section \ref{sec:syst_unc}) of where this disagreement between estimates for the extinction derived from the ionised gas and resolved stellar populations originates, the factor of $\sim 3$ jump in $A_V$ between our single-spaxel (IFU; gaseous) and P2 (photometric; stellar) extinctions is significant. Evidence in support of a lower (< 6000 K) temperature environment includes: the lack of bright H \textsc{ii} regions within an $\sim400$ pc radius of the SN site; limited emission lines in the MUSE spectra (e.g. [O I] $\lambda5577$ and [N II] $\lambda5755$ are not detected); low flux levels in spectra; relatively low number of bright ionising stars; lack of large stellar clusters or star forming complexes; and a dark dust lane intersecting the environment. This could direct us away from measured \citep{1967ApJ...150..825P, 1981AJ.....86..989D, 1987A&A...171..261C} or assumed \citep{2016MNRAS.455.4087G} typical values of $T_e \gtrsim 6000$ K for Galactic and extragalactic H \textsc{ii} regions. Further complicating this case is our lack of constraint over the electron density ($n_e$) of the ionised gas. \citet{2006ApJ...652..401M} and \citet{2024MNRAS.532.2016M} find higher $T_e$ with decreasing $n_e$ for diffuse ionised gas in the ISM of spiral galaxies.  

\subsubsection{Interpreting the 3D Structure and iPTF13bvn’s Extinction}

\citet{2021MNRAS.504.2253S} gave an indepth interpretation of the environments of SNe 2004dg and 2012P, which also show a factor of $\sim 2-3$ increase when looking at the ionised gas extinctions compared to the mean stellar population extinctions. Their schematic diagram attributes this to stellar groups with lower (higher) mean population extinctions being situated in the foreground (background) relative to the observer. Moreover, the highest extinctions that are observed in the ionised gas would place these gaseous structures behind any resolved stellar populations. \citet{2021MNRAS.504.2253S} also acknowledged that this distance interpretation can only be applied to mean population extinctions, but not on the scale of single stars where this interpretation breaks down and local variation is observed in the ISM. When finding the mean population extinctions within our Bayesian approach, we account for this local variation by fitting for the dispersion hyperparameter ($\delta A_V$; see Section \ref{sec:bayesresults}). Transforming these back to a linear scale ($\sigma_{A_{V,k}}=0.058\times10^{\delta A_{V,k}}$), we find $\sigma_{A_{V,1}}=0.35\pm0.04$ mag and $\sigma_{A_{V,2}}=0.10^{+0.06}_{-0.04}$ mag. These results imply increased dispersion about our mean population extinctions for P1 compared to P2, perhaps as a result of seeing P1 through P2.  

SNe 2004dg and 2012P \citep{2021MNRAS.504.2253S} also occurred in NGC 5806, but in the denser central region of another spiral arm and either within (2012P) or nearby ($\sim200$ pc; 2004dg) a bright star-forming complex. Within this complex, the ongoing star formation and young stellar populations clear (in the direction of the observer) a ‘bubble’ in the gas that accumulates in the traversing density wave of the spiral arm. The ionised gas observed in the background is compressed and potentially triggers further star formation.  

Within the context of the environment of iPTF13bvn, the same interpretation can be applied. P2 ($A_V = 0.53$ mag) would sit in the foreground (along our LoS), followed by P1 ($A_V = 0.95$ mag) and then a background source of ionised gas. Yet, unlike SNe 2004dg and 2012P, iPTF13bvn is situated on the edge of the spiral arm. Given that region D and ``13bvn-inner" share similarly positioned environments, them having $\sigma_{\bar{A}_{V}}$ values 2-4 times that of the remaining comparison environments could suggest increased variations in extinction. This could be increased variation in extinction from local (foreground) sources, background sources or a combination of both; the source of increased variation could originate from a loss of a smoothly-varying ionised gas background, replaced by clumpy gas and dust formations. Given that we see a sharp drop-off due to a dust lane in Figure \ref{fig:hst_resampled} (in particular, the F225W images), the background source of compressed photo-ionised gas proposed by \citet{2021MNRAS.504.2253S} seems like a less likely scenario on the edge of a spiral arm (iPTF13bvn). 

Looking to the single spaxel of iPTF13bvn, regardless of constraints over the gas temperature, we only see disagreement between the MUSE vs. HST derived extinctions. With the arguments laid out above, any contribution from local variations to the single spaxel extinction of iPTF13bvn could imply a significantly higher extinction than the value we previously assigned (P2; $A_V = 0.53$ mag). Although this is not definitive, it does still suggest that smoothed population parameters (such as extinction) are potentially underestimating the true values for individual stars and misrepresent the implied progenitor properties. Furthermore, the inability of our ‘13bvn\_outer’ (Figure \ref{fig:muse}) $A_{V, \textrm{env}}$ to be recreated by $\bar{A}_{V}$ and comparatively higher levels of dispersion ($\sigma_{\bar{A}_{V}}$) about the weighted mean extinction (seen for both ‘13bvn\_inner’ and Region D) suggests certain environments (i.e. the edges of spiral arms) are poor environments to assume smooth singular values for population parameters.

\subsubsection{Association with nearby H \textsc{ii} Regions?}

As presented by \citet{2013MNRAS.428.1927C}, Type II SNe are more weakly associated with H \textsc{ii} regions than Type Ib/c. This has been interpreted as typical Type Ib/c progenitors being higher in mass than Type II, but there are past examples of both showing no association to any nearby H \textsc{ii} regions. For iPTF13bvn, our study finds no strong association between the SN site and nearby H \textsc{ii} regions (Table \ref{tab:extinctions} and Figure \ref{fig:muse}), which further justifies the choice of P2 being its natal stellar population. Since we still have no diagnostic for choosing a natal population when more than one is resolved in a given SN environment, future studies may want to use H \textsc{ii} region association as the indicator (i.e. choosing the oldest population when no association is found). Finally, the lack of association between iPTF13bvn and any bright H \textsc{ii} region highlights the danger of the ‘nearest H \textsc{ii} region’ approach used in past environmental studies. In this case, using the age estimates of the nearest H \textsc{ii} regions would lead to an overestimation of the progenitor mass.

Following the scenarios for SNe association with H \textsc{ii} regions by \citet{2013MNRAS.428.1927C} (Section 2.3), iPTF13bvn would fall under Class 2 or 3. As required by Class 2, iPTF13bvn is {\it not} coincident with a detectable star cluster and a H \textsc{ii} region is {\it absent}, favouring an older birth cluster and therefore lower mass progenitor ($< 20\, \textrm{M}_\odot$). We cannot rule out the scenario of a dynamically ejected progenitor (Class 3), given a minimum distance of $\sim 400$ pc from the nearest H \textsc{ii} region to the site of iPTF13bvn. For a typical ejected velocity of 50 km$\textrm{s}^{-1}$, the progenitor would require a lifetime $\gtrsim 8$ Myr and corresponding initial mass $\lesssim 23\, \textrm{M}_\odot$ to traverse this distance. Both 'cases' again align with our P2 result for iPTF13bvn. In active SF galaxies like NGC 5806, H$\alpha$ emission and UV trace SF one-to-one \citep{2007ApJS..173..267S}. It then follows that observing bright UV emission across the site of iPTF13bvn (and broader spiral arm; Section \ref{sec:resampleresults}), but a lack of H \textsc{ii} region association, indicates lower levels of older ($< 100$ Myr) continuous SF, more aligned with the explosion of a lower mass progenitor.  

Probing stellar population ages using the instantaneous starburst model (Section \ref{sec:starburst99}) for the H \textsc{ii} regions surrounding iPTF13bvn creates viable estimates for {\it their} individual stellar populations. By retaining high spatial resolution in our MUSE data, we have shown that iPTF13bvn cannot be conclusively linked to any of these H \textsc{ii} regions and presuming association to the nearest bright H \textsc{ii} region would miss the two unique stellar populations that we observe in Section \ref{sec:bayesresults}. If the case of the environment of iPTF13bvn were to repeat in the sites of other CCSNe, then studies that apply `nearest H \textsc{ii} region' \citep{1992AJ....103.1788V, 1994PASP..106.1276B, 1996AJ....111.2017V}, Voronoi tesselation \citep{2018MNRAS.473.1359L, 2022MNRAS.510.3701S} or NCR pixel statistics \citep{2022MNRAS.513.3564R, 2018A&A...613A..35K, 2008MNRAS.390.1527A} approaches may swallow-up this spatial information into single pixels or apertures in an attempt to improve the data quality when deriving natal population properties. Although improving the data quality (i.e. spectral SNR) this way is good practice, it can remove finer details that would otherwise reveal a more complex environment at the sites of CCSNe capable of assisting in breaking the degeneracy when high and low mass population progenitors are simultaneously observed.

\subsection{Context within other Type Ib Supernovae}

Well observed and studied Type Ib supernovae are still rare objects, slowing our investigation into the potentially shared or segregating properties between SESNe and Type II SNe. A question at the forefront of SESNe research is the fractional contribution of single star vs. binary system progenitors. An investigation by \citet{2012A&A...544L..11Y} suggests that progenitor bolometric luminosities do not provide enough information to probe the single-binary partition. The limited number of highly studied Type Ib SNe are inclined towards binary interaction as the mass-loss mechanism, compared to the strong-winds single massive star model. 

SN 2019yvr \citep{2022MNRAS.510.3701S}, SNe 2016bau and 2015ap \citep{2021MNRAS.505.2530A} and SN 2014C \citep{2020MNRAS.497.5118S} all have identified or suspected lower mass (M$_\textrm{ZAMS}\lesssim 30\, \textrm{M}_\odot$) binary progenitors. An environmental analysis of SN 2019yvr by \citet{2022MNRAS.510.3701S} found three temporally separated episodes of star formation within an active star-forming region. As we find for iPTF13bvn, SN 2019yvr is spatially offset from the stellar surface density and gaseous nebular emission. A low ejecta mass suggests SN 2019yvr belongs to their oldest stellar population, corresponding to a progenitor initial mass of $10.4^{+1.5}_{-1.3}\, \textrm{M}_\odot$. \citet{2021MNRAS.505.2530A} find that both SN 2016bau and 2015ap are best fit by a $12\, \textrm{M}_\odot$ model, too low to be explained by single star evolution. Through SED fitting, \citet{2020MNRAS.497.5118S} find a cluster age of $20.0^{+3.5}_{-2.6}$ Myr for SN 2014C, and binary stellar evolution models estimate a progenitor mass of $\sim 11 \textrm{M}_\odot$. SN 2014C was, however, a peculiar case having transitioned from a Type Ib to IIn classification. Alternatively, \citet{2025A&A...693A..13S} recently studied the progenitor of SN 2019odp. Pseudo-bolometric light curve modeling found an ejecta mass of $M_\textrm{ej} \sim 4-7\, \textrm{M}_\odot$ and nebular spectroscopy modelling found an oxygen mass $> 0.5\, \textrm{M}_\odot$, implying a pre-explosion mass $> 18\, \textrm{M}_\odot$ that best resembles a single massive star, likely a Wolf-Rayet progenitor. Our progenitor mass estimate for iPTF13bvn ($M_{initial,P2} = 20.0\, \textrm{M}_\odot$) substantiates its binary classification and amplifies the case that the majority of local Type Ib SNe are stripped through binary interaction. Although a $20.0\, \textrm{M}_\odot$ single star is potentially explodable \citep{2020MNRAS.492.2578S}, it would lack sufficient mass-loss (via strong-winds) across its lifetime to result in a Type Ib SN. \citet{2021ApJ...922...55B} find that single stars with initial masses $< 30\, \textrm{M}_\odot$ will \textit{not} explode as H-poor SNe, which validates why iPTF13bvn was not observed as a Type IIn or transitional Ibn SN.

Extinctions for Type Ib SNe progenitors encompass a broad range of values that go beyond our relatively high value ($A_{V,P1}=0.95^{+0.07}_{-0.06}$ mag) for the youngest stellar population that we resolve in the environment of iPTF13bvn. \citet{2021MNRAS.505.2530A} found a relatively high extinction of $E(B-V)_{\textrm{host}}=0.566 \pm 0.046$ for SN 2015ap, and similarly \citet{2021MNRAS.504.2073K} find $E(B-V)=0.51^{+0.27}_{-0.16}$ for the site of SN 2019yvr, suggesting our P1 value is \textit{not} extreme for Type Ib SNe.

\section{CONCLUSIONS}
\label{sec:conclusion}
We have examined the environment of the Type Ib SN iPTF13bvn with archival photometric (HST/WFC3) and IFU spectroscopic (VLT/MUSE) data. Using a Bayesian approach to fitting stellar populations to all detected stars within 300 pc of iPTF13bvn produced updated estimates of the SN progenitor age, mass and extinction. Although two unique stellar populations are found, we find the greatest agreement between past progenitor constraints for iPTF13bvn and our oldest resolved stellar population: age $\sim9.33^{+1.38}_{-1.20}$ million years, a corresponding initial mass $M_{initial,P2} = 20.0\, \textrm{M}_\odot$, and extinction $A_{V,P2}=0.53^{+0.10}_{-0.08}$ mag. Spectroscopically examining the same environment and beyond, we find a minimum factor of $\sim 2.5$ increase in extinction when derived from the ionised gas, potentially higher depending on the environmental temperature. However, various signatures (lack of nebular lines, sparse population of massive stars, no associated bright H \textsc{ii} region) point towards a lower temperature ($\leq 6000$ K) ionised environment. Comparing extinction results between individual MUSE-IFU spaxel distributions and the summed spectrum of spaxels within a given aperture, we find evidence for increased variability in extinction in the environment of iPTF13bvn, compared to the central, smoothly-varying regions of the extragalactic spiral arm. The same variability is found in an equivalent comparison environment, with both environments falling on the edge of the spiral arm of the host galaxy NGC 5806. Although this result cannot be definitively interpreted, it suggests that previously used assumptions in the environmental analysis of CCSNe (i.e. Voronoi tessellation in low signal-to-noise environments, using single stellar population properties to constrain progenitors) may fail in environments like iPTF13bvn’s and any similar. Future work expanding this investigation to a catalogue of nearby CCSNe with high-quality photometric and IFU spectroscopic data will reveal if our findings for iPTF13bvn are replicated in other SN sites and if they are connected to their environmental properties, SN classification or stand alone. 

\section*{Acknowledgements}

This work was supported by a faculty scholarship from the University of Sheffield, School of Mathematical and Physical Sciences. This research is based on observations made with the NASA/ESA Hubble Space Telescope obtained from the Space Telescope Science Institute, which is operated by the Association of Universities for Research in Astronomy, Inc., under NASA contract NAS 5–26555. These observations are associated with programs: 15152, 13822 and 12888. This research has made use of the NASA/IPAC Extragalactic Database (NED), which is funded by the National Aeronautics and Space Administration and operated by the California Institute of Technology. Based on observations collected at the European Southern Observatory under ESO programme 097.B-0165. 

\section*{Data Availability}

The data used in this paper is all publicly available through the ESO and STSci archives. Programs 15152, 13822 and 12888 can be found on \url{https://archive.stsci.edu/missions-and-data/hst}. Program 097.B-0165 can be found on \url{http://archive.eso.org}.



\bibliographystyle{mnras}
\bibliography{example} 




\appendix

\section{Details of Spectral Fitting Algorithm}
\label{sec:appendix}

\begin{figure*}
\centering
\includegraphics[width=175mm]{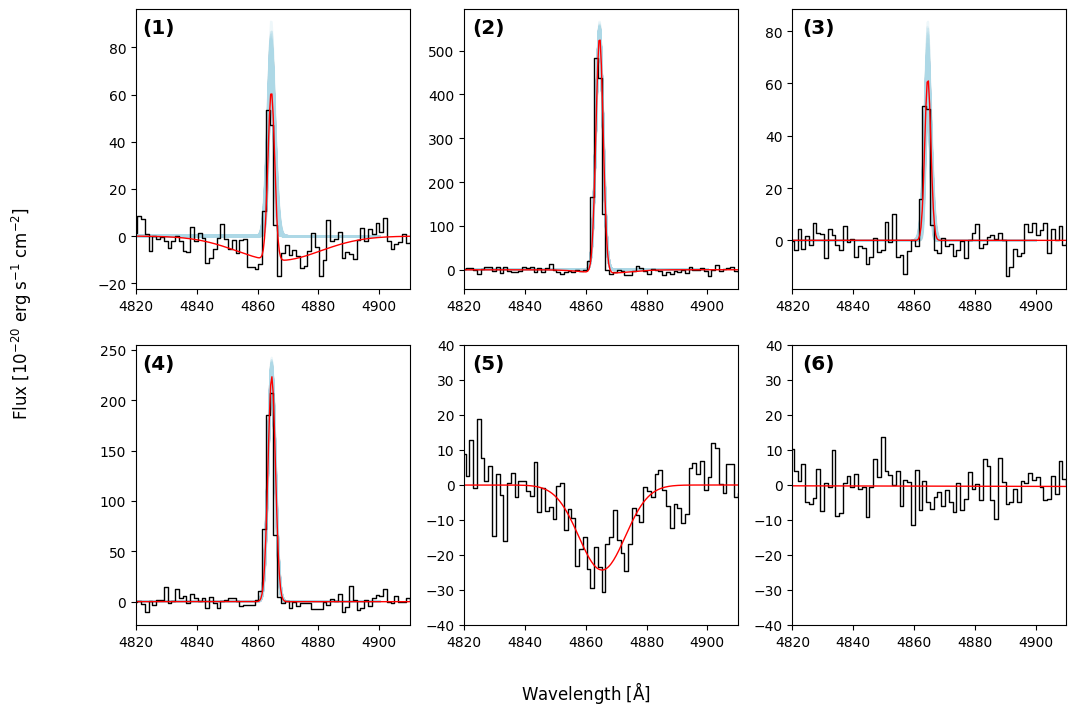}
\vspace{-2mm}
\caption{Examples of fitting algorithm performance across the H$\beta$ line range, with our $\textbf{(1)}\rightarrow\textbf{(6)}$ function numeric labelled in the top right of each subplot (Section \ref{sec:algorithm}). Black lines correspond to the MUSE data, red lines show the overall function fit and shaded blue regions show the error range of the emission line Gaussian for the relevant cases.} 
\label{fig:example_spec}
\end{figure*}

An example of each function (with a corresponding $1\rightarrow6$ function numeric) for our fitting algorithm is presented in Figure \ref{fig:example_spec}. All examples are of the H$\beta$ line. When fitting Gaussians to emission features, the central wavelength value is allowed to vary within the boundary of $\pm 3$ \r{A} to account for the observed differential rotational velocities across the extracted MUSE datacube described in Section \ref{sec:MUSE}.

When fitting a Gaussian to any observed absorption features, we acknowledge that this is a simplified approach that does \textit{not} represent the underlying physics. In reality complex line blending and blurring factors are present, however we are limited in our ability to detect this due to the low spectral resolution of the MUSE data. A more sophisticated approach could fit for the underlying stellar continuum and subtract off to reproduce pure nebular features. Our motivation for not doing so stems from the assumptions that have to be made about the population of stars comprising the stellar continuum. This introduces more model parameters that could accurately constrain the stellar continuum, but poorly recreate the observed absorption. There is also the factor that this approach is easier where there are collections of bright stars to constrain the stellar population, which is not the case across the MUSE datacube. Assumptions about which stars are grouped into which stellar population model are often unjustified. Given that our main goal is to remove any absorption features to better estimate the emission integrated flux (and subsequent extinction estimates), a simple Gaussian seems adequate. In the cases of the 'double Gaussian', \textbf{(1)} and \textbf{(2)} in Section \ref{sec:algorithm}, we only prescribe an absorption fitting when a significant improvement to the reduced-chi-squared is observed. The pure absorption fitting, \textbf{(5)}, described in Section \ref{sec:algorithm} is only utilised to show us where in the MUSE maps we observe stellar populations lacking in nebular emission. 

\bsp	
\label{lastpage}
\end{document}